\documentclass[aps, prd, preprint, showpacs, nofootinbib, floatfix, superscriptaddress]{revtex4-2}
\usepackage{amsmath, amssymb, mathtools, bm, bbm, microtype, tocbibind, tabularx, subfigure, graphicx, placeins}
\allowdisplaybreaks[0]
\usepackage[dvipsnames]{xcolor}
\usepackage{hyperref}
\usepackage{youngtab}\Yboxdim 6.5pt\Ylinethick 0.4pt

\graphicspath{{Figures/}}

\newcommand{\Abs}[1]{\left\lvert#1\right\rvert}
\newcommand{\RealPart}[1]{{\rm Re}(#1)}
\newcommand{\ImaginaryPart}[1]{{\rm Im}(#1)}

\begin{document}

\title{Three-family supersymmetric Pati-Salam models with symplectic groups from intersecting D6-branes}

\author{Adeel Mansha}
\email{adeelmansha@zjnu.edu.cn}
\affiliation{Department of Physics, Zhejiang Normal University, Jinhua 321004, P. R. China}
\affiliation{Zhejiang Institute of Photoelectronics, Zhejiang Normal University, Jinhua 321004, P. R. China}
\affiliation{Zhejiang Institute for Advanced Light Source, Zhejiang Normal University, Jinhua 321004, P. R. China}

\author{Tianjun Li}
\email{tli@itp.ac.cn}
\affiliation{CAS Key Laboratory of Theoretical Physics, Institute of Theoretical Physics, Chinese Academy of Sciences, Beijing 100190, P. R. China}
\affiliation{School of Physical Sciences, University of Chinese Academy of Sciences, Beijing, P. R. China}

\author{Mudassar Sabir}
\thanks{Corresponding author}
\email{mudassar.sabir@uestc.edu.cn}
\affiliation{School of Physics, University of Electronic Science and Technology of China, 
Chengdu, Sichuan 611731, P. R. China}

\author{Lina Wu}
\email{wulina@xatu.edu.cn}
\affiliation{School of Sciences, Xi'an Technological University, Xi'an, Shaanxi 710021, P. R. China}


\begin{abstract}
We construct new three-family ${\cal N}=1$ supersymmetric Pati-Salam models from intersecting D6-branes with original gauge group ${\rm U}(4)_C \times {\rm USp}(2)_L \times {\rm U}(2)_R$ on a Type IIA $\mathbb{T}^6/(\mathbb{Z}_2\times \mathbb{Z}_2)$ orientifold. We find that replacing a ${\rm U}(2)$ with a ${\rm USp}(2)$ group severely restricts the number of three-generation supersymmetric models such that there are only five inequivalent models. Exchanging the left and right sectors, we obtain five dual models with gauge group ${\rm U}(4)_C \times {\rm U}(2)_L \times {\rm USp}(2)_R$. These ten models have different gauge coupling relations at string scale. The highest wrapping number is 4, and one of the models contains no filler O6-planes. Moreover, we discuss in detail the particle spectra, the composite particles through strong coupling dynamics, and the exotic particle decouplings. Also, we study how to realize the string-scale gauge coupling relations in some models.
\end{abstract}

\maketitle
\newpage
\section{Introduction}\label{sec:Intro}

String theory phenomenology is a top-down approach to understand the known particle physics standard model (SM) from a certain Calabi-Yau compactification of extra dimensions. Intersecting D6-branes in type IIA string theory provide a geometric framework to generate the gauge structures, the chiral spectrum and various couplings. For example, the four-dimensional gauge couplings depend on the volume of the cycles wrapped by the D-branes and the gravitational coupling is determined by their total internal volume. Similarly, the cubic Yukawa couplings depend exponentially on the triangular areas of open worldsheet intersections. The general flavor structure and selection rules for intersecting D-brane models has been investigated in \cite{Chamoun:2003pf, Higaki:2005ie}.

Fermions in the standard model (SM) correspond to the chiral representations of the gauge group ${\rm SU}(3)_C\times {\rm SU}(2)_W \times {\rm U}(1)_Y$ such that all gauge anomalies are canceled. Simply placing parallel D-branes in flat space does not yield chiral fermions. Instead, to realize the chiral fermions we need to either place D-branes on orbifold singularities \cite{Aldazabal:2000sa} or else consider intersecting D-branes on generalized orbifolds called orientifolds \cite{Shiu:1998pa, Cvetic:2001tj}. In orientifolds, both the discrete internal symmetries of the world-sheet theory and the products of internal symmetries with world-sheet parity reversal become gauged.

Several supersymmetric Pati-Salam models from intersecting D6-branes on a $\mathbbm{T}^6/(\mathbbm{Z}_2\times \mathbbm{Z}_2)$ orientifold from IIA string theory were constructed in ref.~\cite{Cvetic:2004ui} up to the wrapping number 3. Recently, utilizing machine learning algorithm in ref.~\cite{Li:2019nvi} several new models up to the wrapping number 5 were constructed. The phenomenology of models up to the wrapping number of 3 was first studied in refs.~\cite{Chen:2007px, Chen:2007zu} while the phenomenology of a newly found model with a wrapping number 5 was explored in \cite{Sabir:2022hko}. In ref.~\cite{He:2021kbj, He:2021gug}, a mathematical search algorithm was used to possibly obtain the complete landscape of supersymmetric Pati-Salam models, where again the highest wrapping number turned out to be 5 as already indicated from the random search. However, an interesting variation of the Pati-Salam gauge group that was not taken into account in previous searches \cite{Cvetic:2004ui, Li:2019nvi, He:2021kbj, He:2021gug, Li:2021pxo} is the ${\rm SU}(4)_C \times {\rm USp}(2)_L \times {\rm SU}(2)_R$ and ${\rm SU}(4)_C \times {\rm USp}(2)_L \times {\rm USp}(2)_R$ where either one or both of the usual SU(2)s are replaced by USp(2) groups. The construction of USp(2) groups can be readily achieved by taking the corresponding stacks of D6-branes parallel to any of the O6-planes or their $\mathbbm{Z}_2$ images as exemplified in ref.~\cite{Cvetic:2004nk}.

In this paper, we present results for the search of $\mathcal{N}=1$ supersymmetric Pati-Salam models with the gauge group ${\rm SU}(4)_C \times {\rm USp}(2)_L \times {\rm SU}(2)_R$ in intersecting D-branes for the first time. Unlike \cite{Cvetic:2004nk}, we replace only one of the unitary gauge group with the symplectic group. Unlike the U(2) gauge group, the USp(2) group is simpler to deal with because there is no associated global anomalous U(1) group which is generic in the unitary groups. We have found 53 models out of which there are only five inequivalent models. By swapping the left and the right sectors yields dual models with the gauge group ${\rm SU}(4)_C \times {\rm SU}(2)_L \times {\rm USp}(2)_R$ having completely different low energy spectra and new gauge coupling relations. Thus, we can have ten different gauge coupling unification relations taking into account the exchange of the left and right sectors. The highest wrapping number was found to be 4 and one of the models contains no filler O6-planes. We detail out the complete particle spectra and also discuss the decoupling of exotic particles through strong coupling dynamics.

The plan of the paper is as follows. In section~\ref{sec:orientifold}, we review the model building rules for $\mathcal{N}=1$ supersymmetric Pati-Salam models with the gauge group ${\rm SU}(4)_C \times {\rm USp}(2)_L \times {\rm SU}(2)_R$ on a $\mathbbm{T}^6/(\mathbbm{Z}_2\times \mathbbm{Z}_2)$ orientifold. We discuss the gauge symmetry breaking via the brane-splitting mechanism to yield the standard model gauge group. In section~\ref{sec:Pheno} we discuss the salient phenomenological features and elaborate upon the decoupling of the various exotic particles from the low energy spectra. In section~\ref{sec:g-c-u} we discuss the gauge coupling relations and display the results of two-loop RGE running by adding vector-like particles to achieve string-scale gauge coupling relations at  $5\times10^{17}$ GeV. Finally, we conclude in section~\ref{sec:conclusion} by summarizing the main findings and the limitations of the phenomenologically interesting models.

\section{The Pati-Salam model building from $\mathbb{T}^6/(\mathbb{Z}_2\times \mathbb{Z}_2)$ orientifold} \label{sec:orientifold}

In the orientifold $\mathbbm{T}^6/(\mathbbm{Z}_2\times \mathbbm{Z}_2)$, $\mathbbm{T}^6$ is a factorizable product of three two-tori with the orbifold group $(\mathbbm{Z}_2\times \mathbbm{Z}_2)$ having the generators $\theta$ and $\omega$ which are respectively associated with the twist vectors $(1/2,-1/2,0)$ and $(0,1/2,-1/2)$ such that their action on complex coordinates $z_i$ is given by,
\begin{align}
\theta: \quad (z_1,z_2,z_3) &\to (-z_1,-z_2,z_3), \nonumber \\
\omega: \quad (z_1,z_2,z_3) &\to (z_1,-z_2,-z_3). \label{orbifold}
\end{align}
Orientifold projection is the gauged $\Omega {\cal R}$ symmetry, where $\Omega$ is world-sheet parity that interchanges the left- and right-moving sectors of a closed string and swaps the two ends of an open string as,
\begin{align}
\textrm{Closed}:  &\quad \Omega : (\sigma_1, \sigma_2) \mapsto (2\pi -\sigma_1, \sigma_2), \nonumber \\
\textrm{Open}:  &\quad  \Omega : (\tau, \sigma) \mapsto (\tau, \pi - \sigma) ,
\end{align}
and ${\cal R}$ acts as complex conjugation on coordinates $z_i$. This results in four different kinds of orientifold 6-planes (O6-planes) corresponding to $\Omega {\cal R}$, $\Omega {\cal R}\theta$, $\Omega {\cal R}\omega$, and $\Omega {\cal R}\theta\omega$ respectively. These orientifold projections are only consistent with either the rectangular or the tilted complex structures of the factorized two-tori. Denoting the wrapping numbers for the rectangular and tilted tori as $n_a^i[a_i]+m_a^i[b_i]$ and $n_a^i[a'_i]+m_a^i[b_i]$ respectively, where $[a_i']=[a_i]+\frac{1}{2}[b_i]$. Then a generic 1-cycle $(n_a^i,l_a^i)$ satisfies $l_{a}^{i}\equiv m_{a}^{i}$ for the rectangular two-torus and $l_{a}^{i}\equiv 2\tilde{m}_{a}^{i}=2m_{a}^{i}+n_{a}^{i}$ for the tilted two-torus such that $l_a^i-n_a^i$ is even for the tilted tori.

The homology cycles for a stack $a$ of $N_a$ D6-branes along the cycle $(n_a^i,l_a^i)$ and their $\Omega {\cal R}$ images ${a'}$ stack of $N_a$ D6-branes with cycles $(n_a^i,-l_a^i)$ are respectively given as,
\begin{align}
[\Pi_a ]&=\prod_{i=1}^{3}\left(n_{a}^{i}[a_i]+2^{-\beta_i}l_{a}^{i}[b_i]\right), \nonumber \\
[\Pi_{a'}] &=\prod_{i=1}^{3}\left(n_{a}^{i}[a_i]-2^{-\beta_i}l_{a}^{i}[b_i]\right),
\end{align}
where $\beta_i=0$ or $\beta_i=1$ for the rectangular or tilted $i^{\rm th}$ two-torus, respectively.
The homology three-cycles, which are wrapped by the four O6-planes, are given by
\begin{alignat}{2}
\Omega {\cal R} :            &\quad &        [\Pi_{\Omega {\cal R}}] &= 2^3 [a_1]\times[a_2]\times[a_3],  \nonumber\\
\Omega {\cal R}\omega :      &&        [\Pi_{\Omega {\cal R}\omega}] &=-2^{3-\beta_2-\beta_3}[a_1]\times[b_2]\times[b_3],  \nonumber\\
\Omega {\cal R}\theta\omega : && [\Pi_{\Omega {\cal R}\theta\omega}] &=-2^{3-\beta_1-\beta_3}[b_1]\times[a_2]\times[b_3], \nonumber\\
\Omega {\cal R}\theta :      &&       [\Pi_{\Omega {\cal R} \theta}] &=-2^{3-\beta_1-\beta_2}[b_1]\times[b_2]\times[a_3]. \label{orienticycles}
\end{alignat}
The intersection numbers can be calculated in terms of wrapping numbers as,
\begin{align}
I_{ab} &=[\Pi_a][\Pi_b] =2^{-k}\prod_{i=1}^3(n_a^il_b^i-n_b^il_a^i),\nonumber\\
I_{ab'}&=[\Pi_a]\left[\Pi_{b'}\right] =-2^{-k}\prod_{i=1}^3(n_{a}^il_b^i+n_b^il_a^i),\nonumber\\
I_{aa'}&=[\Pi_a]\left[\Pi_{a'}\right] =-2^{3-k}\prod_{i=1}^3(n_a^il_a^i),\nonumber\\
I_{aO6}&=[\Pi_a][\Pi_{\rm O6}] =2^{3-k}(-l_a^1l_a^2l_a^3+l_a^1n_a^2n_a^3+n_a^1l_a^2n_a^3+n_a^1n_a^2l_a^3),\label{intersections}
\end{align}
where $k=\sum_{i=1}^3\beta_i$ and $[\Pi_{\rm O6}]=[\Pi_{\Omega {\cal R}}]+[\Pi_{\Omega {\cal R}\omega}]+[\Pi_{\Omega {\cal R}\theta\omega}]+[\Pi_{\Omega {\cal R}\theta}]$.

\subsection{Constraints from tadpole cancellation and supersymmetry}\label{subsec:constraints}

Since D6-branes and O6-orientifold planes are the sources of Ramond-Ramond charges they are constrained by the Gauss's law in compact space implying that the sum of D-brane and cross-cap RR-charges must vanish \cite{Gimon:1996rq}
\begin{eqnarray}\label{RRtadpole}
\sum_a N_a \left[\Pi_a+\Pi_{a'}\right]-4[\Pi_{\rm O6}]=0,
\end{eqnarray}
where the second term arises from the O6-planes, which have $-4$ RR charges in D6-brane charge units. RR tadpole constraint is sufficient to cancel the ${\rm SU}(N_a)^3$ cubic non-Abelian anomaly while U(1) mixed gauge and gravitational anomaly or $[{\rm SU}(N_a)]^2 {\rm U}(1)$ gauge anomaly can be cancelled by the Green-Schwarz mechanism, mediated by untwisted RR fields \cite{Green:1984sg}.

Let us define the following products of wrapping numbers,
\begin{alignat}{4}
A_a &\equiv -n_a^1n_a^2n_a^3, &\quad B_a &\equiv n_a^1l_a^2l_a^3, &\quad C_a &\equiv l_a^1n_a^2l_a^3, \quad & D_a &\equiv l_a^1l_a^2n_a^3, \nonumber\\
\tilde{A}_a &\equiv -l_a^1l_a^2l_a^3, & \tilde{B}_a &\equiv l_a^1n_a^2n_a^3, & \tilde{C}_a &\equiv n_a^1l_a^2n_a^3, & \tilde{D}_a &\equiv n_a^1n_a^2l_a^3.\,\label{variables}
\end{alignat}
Cancellation of RR tadpoles requires introducing a number of orientifold planes also called ``filler branes'' that trivially satisfy the four-dimensional ${\cal N}=1$ supersymmetry conditions. The no-tadpole condition is given as,
\begin{align}\label{notadpoles}
 -2^k N^{(1)}+\sum_a N_a A_a &= -2^k N^{(2)}+\sum_a N_a B_a = \nonumber\\
 -2^k N^{(3)}+\sum_a N_a C_a &= -2^k N^{(4)}+\sum_a N_a D_a = -16,\,
\end{align}
where $2 N^{(i)}$ is the number of filler branes wrapping along the $i^{\rm th}$ O6-plane. The filler branes belong to the hidden sector USp group and carry the same wrapping numbers as one of the O6-planes as shown in table~\ref{orientifold}. USp group is hence referred with respect to the non-zero $A$, $B$, $C$ or $D$-type.

\begin{table}[htb]
\renewcommand{\arraystretch}{1.3} 
\caption{The wrapping numbers for four O6-planes.} \label{orientifold} 
\begin{tabular}{c}$\begin{array}{|c|c|c|}
\hline
\text{Orientifold action} & \text{O6-plane} & (n^1,l^1)\times (n^2,l^2)\times (n^3,l^3)\\
\hline\hline
    \Omega {\cal R}& 1 & (2^{\beta_1},0)\times (2^{\beta_2},0)\times (2^{\beta_3},0) \\
\hline
    \Omega {\cal R}\omega& 2& (2^{\beta_1},0)\times (0,-2^{\beta_2})\times (0,2^{\beta_3}) \\
\hline
    \Omega {\cal R}\theta\omega& 3 & (0,-2^{\beta_1})\times (2^{\beta_2},0)\times (0,2^{\beta_3}) \\
\hline
    \Omega {\cal R}\theta& 4 & (0,-2^{\beta_1})\times (0,2^{\beta_2})\times (2^{\beta_3},0) \\
\hline
\end{array}$ 
\end{tabular}\end{table}

Preserving ${\cal N}=1$ supersymmetry in four dimensions after compactification from ten-dimensions restricts the rotation angle of any D6-brane with respect to the orientifold plane to be an element of ${\rm SU}(3)$, i.e.
\begin{equation}
\theta^a_1 + \theta^a_2 + \theta^a_3 = 0 \mod 2\pi ,
\end{equation}
with $\theta^a_j = \arctan (2^{- \beta_j} \chi_j l^a_j/n^a_j)$. $\theta_i$ is the angle between the D6-brane and orientifold-plane in the $i^{\rm th}$ two-torus and $\chi_i=R^2_i/R^1_i$ are the complex structure moduli for the $i^{\rm th}$ two-torus.
${\cal N}=1$ supersymmetry conditions are given as,
\begin{eqnarray}
x_A\tilde{A}_a+x_B\tilde{B}_a+x_C\tilde{C}_a+x_D\tilde{D}_a=0,\nonumber\\
\frac{A_a}{x_A}+\frac{B_a}{x_B}+\frac{C_a}{x_C}+\frac{D_a}{x_D} < 0, \label{susyconditions}
\end{eqnarray}
where $x_A=\lambda,\; x_B=2^{\beta_2+\beta_3}\cdot\lambda /\chi_2\chi_3,\; x_C=2^{\beta_1+\beta_3}\cdot\lambda /\chi_1\chi_3,\; x_D=2^{\beta_1+\beta_2}\cdot\lambda /\chi_1\chi_2$.

Orientifolds also have discrete D-brane RR charges classified by the $\mathbbm{Z}_2$ K-theory groups, which are subtle and invisible by the ordinary homology~\cite{Witten:1998cd, Cascales:2003zp, Marchesano:2004yq, Marchesano:2004xz}, which should also be taken into account~\cite{Uranga:2000xp}. The K-theory conditions are,
\begin{eqnarray}\label{K-charges}
\sum_a N_a \tilde{A}_a  = \sum_a  N_a  \tilde{B}_a = \sum_a  N_a  \tilde{C}_a = \sum_a  N_a \tilde{D}_a = 0 \mod 4 ~.
\end{eqnarray}
In our case, we avoid the nonvanishing torsion charges by taking an even number of D-branes, {\it i.e.}, $N_a \in 2 \mathbbm{Z}$.

\begin{table}[htb]
\renewcommand{\arraystretch}{1.3}\centering
\caption{General spectrum for intersecting D6-branes at generic angles, where ${\cal M}$ is the multiplicity, and  $a_{\protect\yng(2)}$ and $a_{\protect\yng(1,1)}$ denote respectively the symmetric and antisymmetric representations of ${\rm U}(N_a/2)$. Positive intersection numbers in our convention refer to the left-handed chiral supermultiplets.\\}
\begin{tabular}{c}$\begin{array}{|c|c|}
\hline
\text{\bf Sector} & \phantom{more space inside this box}{\bf Representation} \phantom{more space inside this box} \\
\hline\hline
aa                & {\rm U}(N_a/2) \text{ vector multiplet}  \\
                  & \text{3 adjoint chiral multiplets}  \\
\hline ab+ba      & {\cal M}(\frac{N_a}{2}, \frac{\overline{N_b}}{2})= I_{ab}(\yng(1)_{a},\overline{\yng(1)}_{b}) \\
\hline ab'+b'a    & {\cal M}(\frac{N_a}{2}, \frac{N_b}{2})=I_{ab'}(\yng(1)_{a},\yng(1)_{b}) \\
\hline aa'+a'a    & {\cal M} (a_{\yng(2)})= \frac{1}{2} (I_{aa'} - \frac{1}{2} I_{aO6}) \\
                  & {\cal M} (a_{\yng(1,1)_{}})= \frac{1}{2} (I_{aa'} + \frac{1}{2} I_{aO6})  \\
\hline
\end{array}$
\label{tab:spectrum}
\end{tabular}\end{table}
The general particle representations for intersecting D6-branes models at angles are shown in table \ref{tab:spectrum}. Following the convention of \cite{Cvetic:2004ui} the $N$ number of D6-brane stacks corresponds to U($N/2$) and USp($N$) respectively. A positive intersection number in our convention refers to the left-chiral supermultiplet. To have three families of the SM fermions, we need one torus to be tilted, which is chosen to be the third torus. So we have  $\beta_1=\beta_2=0$ and $\beta_3=1$.

Higgs arise as vector-like pairs from the massless open string states from the $bc$ sector. The anomalies from three global U(1)s of ${\rm U}(4)_C$, ${\rm U}(2)_R$ and ${\rm U}(2)_d$ are cancelled by the Green-Schwarz mechanism, and the gauge fields of these U(1)s obtain masses via the linear $B\wedge F$ couplings. Thus, the effective gauge symmetry is ${\rm SU}(4)_C\times {\rm USp}(2)_L\times {\rm SU}(2)_R$.

\subsection{Gauge symmetry breaking}\label{sec:brane-splitting}

To obtain SM or standard-like models via the mechanism of intersecting D6-branes, there should be at least two extra U(1) gauge symmetries for either supersymmetric models or non-supersymmetric models, as a result of the constraints on the quantum number of the right handed electron \cite{Ibanez:2001nd, Cvetic:2001nr, Cvetic:2002pj, Cvetic:2003xs}. Among these two U(1) gauge symmetries, one is lepton number symmetry ${\rm U}(1)_L$ and another ${\rm U}(1)_{I_{3R}}$ is an analogy for right-hand weak isospin. We have the hypercharge $Q_Y$ expressed in the form
\begin{equation}\label{eq: hypercharge}
    Q_Y = Q_{I_{3R}} + \frac{Q_B - Q_L}{2}.
\end{equation}
The baryonic charge $Q_B$ arises from ${\rm U}(1)_B$, via the decomposition ${\rm U}(3)_C \simeq{\rm SU}(3)_C \times {\rm U}(1)_B$. On the other hand, since the ${\rm U}(1)_{I3R}$ gauge field should be massless, the gauge group ${\rm U}(1)_{I3R}$ must come from the non-abelian component of ${\rm U}(2)_R$ or ${\rm USp}(2)_R$ symmetry, otherwise the ${\rm U}(1)_{I3R}$ will acquire mass from the $B\wedge F$ couplings. To get an anomaly free ${\rm U}(1)_{B-L}$, the ${\rm U}(1)_L$ symmetry should come from some non-abelian group for similar reasons.
In previous studies of supersymmetric model building~\cite{Cvetic:2001nr, Cvetic:2003xs}, ${\rm U}(1)_{I3R}$ comes from $\mathrm{USp}$ groups. These models indeed have two extra anomaly-free U(1) symmetries, and have at least $8$ Higgs doublets. One could in principal break their symmetry groups down to the SM symmetry, but cannot do this without violating the D-flatness and F-flatness, thus the supersymmetry.

For each stack of the stacks $a$, $c$, and $d$ of D6-branes comprising ${\rm U}(4)_C$, ${\rm U}(2)_R$ and ${\rm U}(2)$ gauge groups, there is an overall U(1) gauge symmetry.
The anomalies of the overall U(1) symmetries are canceled by the generalized Green-Schwarz mechanism \cite{Aldazabal:2000dg, Ibanez:2001nd, Cvetic:2001nr}, while their fields get massive from the linear $B\wedge F$ couplings. The gauge symmetry  ${\rm SU}(4)_C \times {\rm USp}(2)_L \times {\rm SU}(2)_R$ can be further broken down to SM gauge symmetry by D6-brane splitting and Higgs mechanism. Firstly, one can split stack $a$ of $N_a = 8$ D6-branes into stack $a_1$ of $N_{a_1}=6$ D6-branes and stack $a_2$ $N_{a_2} = 2$ D6-branes. Then the ${\rm U}(4)_C$ symmetry breaks down to ${\rm U}(3)\times {\rm U}(1)$.

We denote the numbers of symmetric and anti-symmetric representations for ${\rm SU}(4)_C$ , ${\rm USp}(2)_L$ and ${\rm SU}(2)_R$ by $n^a_{\yng(2)}$ and $n^a_{\yng(1,1)}$,  $n^b_{\yng(2)}, n^b_{\yng(1,1)}$ and $n^c_{\yng(2)}, n^c_{\yng(1,1)}$. After splitting, the symmetric and anti-symmetric representations of ${\rm SU}(4)_C$ descend to symmetric representations of  ${\rm SU}(3)_C$ and ${\rm U}(1)_{B-L}$, and anti-symmetric representations of ${\rm SU}(3)_C$. Note that there are $I_{a_1 a'_2}$ new fields, arising from the intersection of $a_1$ stack and $a_2$ stack of D6-branes. The anomaly-free gauge symmetry is ${\rm SU}(3)_C \times {\rm U}(1)_{B-L}$, a subgroup of ${\rm SU}(4)_C$.

Similarly, the  stack $c$ of $N_c = 4$ D6-branes can be broken into stack $c_1$ of $N_{a_1} = 2$ D6-branes and stack $c_2$ $N_{a_2} = 2$ D6-branes. Then the ${\rm U}(2)_R$ symmetry breaks down to ${\rm U}(1)_{I3R}$. The symmetric representations of ${\rm SU}(2)_R$ descend to the symmetric representations of ${\rm U}(1)_{I3R}$ only. Also, there are $I_{c_1 c'_2}$ new fields, arising from the intersection of $c_1$ stack and $c_2$ stack of D6-branes. The anomaly-free gauge symmetry is ${\rm U}(1)_{I3R}$, a subgroup of ${\rm SU}(2)_R$. After splitting, the gauge symmetry of our model breaks down to ${\rm SU}(3)_C \times {\rm USp}(2)_L \times {\rm U}(1)_{B-L}\times {\rm U}(1)_{I3R}$ .

To get just the SM gauge symmetry, we assume the minimal squared distance $Z^2_{a_2 c'_1}$ between $a_2$ stack and the orientifold image of $c_1$ stack of D6-branes to be very small. Then there are $I_{a_2 c'_1}$ chiral multiplets of light fermions, arising from the open string stretching between $a_2$ stack of D6-branes and the orientifold image of $c_1$ stack of D6-branes. These particles break down ${\rm SU}(3)_C \times {\rm USp}(2)_L \times {\rm U}(1)_{B-L}\times {\rm U}(1)_{I3R}$ to the SM gauge symmetry, playing the same role as the right-handed neutrinos and their complex conjugates. Meanwhile, they preserve the D-flatness and F-flatness, thus the supersymmetry. In conclusion, the whole symmetry-breaking chain is
\begin{equation}
    \begin{split}
        {\rm SU}(4)_C \times {\rm USp}(2)_L \times {\rm SU}(2)_R &\xrightarrow{a \to a_1 + a_2} {\rm SU}(3)_C \times {\rm USp}(2)_L \times {\rm SU}(2)_R \times {\rm U}(1)_{B-L}\\
        & \xrightarrow{c\to c_1 + c_2} {\rm SU}(3)_C \times {\rm USp}(2)_L \times {\rm U}(1)_{I3R} \times {\rm U}(1)_{B-L}\\
        & \xrightarrow{\text{Higgs Mechanism}}{\rm SU}(3)_C \times {\rm USp}(2)_L \times {\rm U}(1)_Y\,.
    \end{split}
\end{equation}
The process of dynamical supersymmetry breaking has been studied in \cite{Cvetic:2003yd} for D6-brane models from Type IIA orientifolds.
The complex structure moduli $U$ can be obtained from the supersymmetry conditions as \cite{Sabir:2022hko},
\begin{align}\label{U-moduli}
U^i & = \frac{i R_2^i}{R_1^i+\frac{\beta_i}{2} iR_2^i} = \frac{4i \chi^i+2\beta_i^2\chi_i^2}{4+\beta_i^2\chi_i^2}, \qquad \because \chi^i \equiv \frac{R_2^i}{R_1^i}.
\end{align}
These upper case moduli in string theory basis can be transformed in to lower case {$s$, $t$, $u$} moduli in field theory basis as \cite{Lust:2004cx},
\begin{equation}\label{eq:sugra-string-basis}
    \begin{split}
    \RealPart{s}   &= \frac{e^{-\phi_4}}{2\pi}\frac{\sqrt{\ImaginaryPart{U^1} \,  \ImaginaryPart{U^2}  \, \ImaginaryPart{U^3}  }   }{\Abs{U^1 U^2 U^3}  }\,, \\
    \RealPart{u^1} &= \frac{e^{-\phi_4}}{2\pi} \sqrt{\frac{\ImaginaryPart{U^1}  }{\ImaginaryPart{U^2}\,  \ImaginaryPart{U^3}}  } \Abs{\frac{U^2 U^3}{U^1}}\,, \\
    \RealPart{u^2} &= \frac{e^{-\phi_4}}{2\pi} \sqrt{\frac{\ImaginaryPart{U^2}  }{\ImaginaryPart{U^1}\,  \ImaginaryPart{U^3}}  } \Abs{\frac{U^1 U^3}{U^2}}\,, \\
    \RealPart{u^3} &= \frac{e^{-\phi_4}}{2\pi} \sqrt{\frac{\ImaginaryPart{U^3}  }{\ImaginaryPart{U^1}\,  \ImaginaryPart{U^2}}  } \Abs{\frac{U^1 U^2}{U^3}}\,.
    \end{split}
\end{equation}
where $\phi_4$ is the four dimensional dilaton which is related to the supergravity moduli as
\begin{equation}
2\pi e^{\phi_4}=\Big(\mathrm{Re}(s)\,\mathrm{Re}(u_1)\,\mathrm{Re}(u_2)\,\mathrm{Re}(u_3)\Big)^{-1/4}.
\end{equation}
Note that the three moduli parameter $\chi_1$, $\chi_2$, $\chi_3$ are not independent, as they can be expressed in terms of $x_A, x_B, x_C, x_D$ and the latter parameters are related by the supersymmetric condition~\eqref{susyconditions}. Actually, one can determine $\chi_1, \chi_2, \chi_3$ up to an overall coefficient, namely an action of dilation on these parameters. So one has to stabilize this dilation to determine all the moduli parameters. The holomorphic gauge kinetic function for any D6-brane stack $x$ wrapping a calibrated 3-cycle is given as \cite{Blumenhagen:2006ci},
\begin{equation}
f_x = \frac{1}{2\pi \ell_s^3}\left[e^{-\phi}\int_{\Pi_x} \mbox{Re}(e^{-i\theta_x}\Omega_3)-i\int_{\Pi_x}C_3\right],
\end{equation}
where the integral involving 3-form $\Omega_3$ gives,
\begin{equation}
\int_{\Pi_x}\Omega_3 = \frac{1}{4}\prod_{i=1}^3(n_x^iR_1^i + 2^{-\beta_i}il_x^iR_2^i).
\end{equation}
It can then be shown that,
\begin{align}
f_x &=
\frac{1}{4\kappa_x}(n_x^1\,n_x^2\,n_x^3\,s-\frac{n_x^1\,l_x^2\,l_x^3\,u^1}{2^{(\beta_2+\beta_3)}}-\frac{l_x^1\,n_x^2\,l_x^3\,u^2}{2^{(\beta_1+\beta_3)}}-
\frac{l_x^1\,l_x^2\,n_x^3\,u^3}{2^{(\beta_1+\beta_2)}}),
\label{kingauagefun}
\end{align}
where the factor ${\kappa}_x$ is the Kac-Moody level of the corresponding gauge Kac-Moody algebra associated to the D6-brane stacks such that ${\kappa}_x =1$ for ${\rm U}(N_x)$ and ${\kappa}_x =2$ for USp($2N_x$) or SO($2N_x$) \cite{Ginsparg:1987ee, Hamada:2014eia}.
The gauge coupling constant related to any stack $x$ of D6-branes is
\begin{equation}
    g_x^{-2} = \Abs{\RealPart{f_x}}\,,
\end{equation}
Since, the standard model hypercharge ${\rm U}(1)_Y$ is a linear combination of several ${\rm U}(1)$s,
\begin{equation}
Q_Y=\frac{2}{3}Q_{a}-Q_{c}.
\end{equation}
Therefore, the holomorphic gauge kinetic function for the hypercharge is also taken as a linear combination of the kinetic gauge functions from all of the stacks as \cite{Blumenhagen:2003jy, Ibanez:2001nd},
\begin{equation}\label{fY}
f_Y=\frac{2}{3}f_{a}+f_{c},
\end{equation}
where we have taken into account that our U(1)s are not canonically normalized. And the coupling constant for the hypercharge $g_Y$ can be determined as,
\begin{equation}
g^{-2}_Y = \Abs{\RealPart{f_Y}}\,.
\end{equation}
Hence, the tree-level gauge couplings have the relation,
\begin{equation}
g_a^2 = k_2\, g_b^2 = k_y \left(\frac{5}{3}\right)g_Y^2 = \gamma (\pi e^{-\phi_4}),
\end{equation}
where $g_a$, $g_b$ and $g_Y$ denote the strengths of the strong coupling, the weak coupling, and the hypercharge coupling respectively.
Moreover, the K\"ahler potential takes the following form,
\begin{equation}\label{eq: Kaehler_potential}
    K = - \ln(S + \overline{S}) - \sum_{i=1}^3\ln(U^i + \overline{U}^i).
\end{equation}
And the one-loop beta functions for each $\mathrm{USp}(N^{(i)})$ arising from $2 N^{(i)}$ filler branes are \cite{Cvetic:2004ui},
\begin{equation}\label{eq: beta_functions}
\begin{split}
    \beta^g_i &= - 3\left(\frac{N^{(i)}}{2} + 1\right) + 2 \Abs{I_{ai}} + \Abs{I_{bi}} + \Abs{I_{ci}} + \Abs{I_{di}} + 3\left(\frac{N^{(i)}}{2} - 1\right) \\
    & = -6 + 2\Abs{I_{ai} }+\Abs{I_{bi} } + \Abs{I_{ci} } + \Abs{I_{di}} .
\end{split}
\end{equation}

\section{Phenomenological features and particle spectra}\label{sec:Pheno}

Employing novel random and supervised scanning algorithm \cite{Li:2019nvi} we systematically scanned for the ${\cal N}=1$ supersymmetric Pati-Salam models with the gauge group ${\rm SU}(4)_C \times {\rm USp}(2)_L \times {\rm SU}(2)_R \times {\rm SU}(2)_{\rm hidden} \times {\rm USp}(N)_{\rm hidden}$. We obtained a total of 53 new models where the largest wrapping number was 4. After imposing various type I and type II T-duality equivalence conditions \cite{Cvetic:2004ui}, there only remain five unique models as listed in the appendix~\ref{appendix}. In particular, the Model~\hyperref[model0]{1} has no USp group. Models~\hyperref[model2.1]{2} and~\hyperref[model2.2]{3} each have a ${\rm USp}(2)$ group from the stack 1 and stack 2 respectively. Similarly Model~\hyperref[model4]{4} possesses a ${\rm USp}(4)$ group while Model~\hyperref[model6]{5} contains a ${\rm USp}(6)$ group in the hidden sector.

\begin{table}[htb]
\centering \footnotesize\renewcommand{\arraystretch}{1.3}
\caption{The spectrum of chiral and vector-like superfields, and their quantum numbers under the gauge symmetry ${\rm SU}(4)\times {\rm USp}(2)_L \times {\rm SU}(2)_R \times {\rm SU}(2)$ for the Model~\hyperref[model0]{1}.\\}\label{tab:spec0}
\begin{tabular}{c}$\begin{array}{|c|c||r|r|r||c|c|c|}\hline
\text{Model~\hyperref[model0]{1}} & \text{Quantum Number}  & Q_4 & Q_{2L} & Q_{2R} & Q_{em} & B-L & \text{Field} \\
\hline\hline
ab               & 3 \times (4,\overline{2},1,1)           &  1 & -1 &  0  & -\frac{1}{3}, \frac{2}{3}, -1, 0 & \frac{1}{3}, -1  &  Q_L, L_L\\
ac               & 3\times (\overline{4},1,2,1)            & -1 &  0 &  1  & \frac{1}{3}, -\frac{2}{3}, 1, 0  & -\frac{1}{3}, 1  &  Q_R, L_R\\
bc               & 4 \times (1,\overline{2},2,1)           &  0 & -1 &  1  & 1, 0, 0, -1 &  0  &  H^i_u, H^i_d\\
bc'              & 4\times (1,\overline{2},\overline{2},1) &  0 & -1 & -1  & -1, 0, 0, 1 &  0  &  H'^i\\
ad               & 4\times (4,1,1,\overline{2})            &  1 &  0 &  0  & \frac{1}{6}, -\frac{1}{2} & \frac{1}{3}, -1 &    \\
ad'              & 4\times (4,1,1,2)                       &  1 &  0 &  0  & \frac{1}{6}, -\frac{1}{2} & \frac{1}{3}, -1 &    \\
bd               & 2\times (1,2,1,\overline{2})            &  0 &  1 &  0  & \pm \frac{1}{2} & 0  &    \\
bd'              & 2\times (1,\overline{2},1,\overline{2}) &  0 & -1 &  0  & \mp \frac{1}{2} & 0  &    \\
cd               & 4\times (1,1,2,\overline{2})            &  0 &  0 &  1  & \pm \frac{1}{2} & 0  &    \\
cd'              & 10\times (1,1,\overline{2},\overline{2})&  0 &  0 & -1  & \mp \frac{1}{2} & 0  &    \\
a_{\overline{\yng(2)}}   & 2\times(\overline{10},1,1,1)    & -2 &  0 &  0  & \frac{1}{3}, -\frac{1}{3}, -1  & \frac{2}{3}, -2 &    \\
a_{\yng(1,1)}            & 2\times(6,1,1,1)                &  2 &  0 &  0  & \frac{1}{3}, -1  & \frac{2}{3}, -2 &    \\
c_{\yng(2)}              & 1\times(1,1,3,1)                &  0 &  0 &  2  & 0, \pm 1  &  0 &    \\
c_{\overline{\yng(1,1)}} & 1\times(1,1,\overline{1},1)     &  0 &  0 & -2  & 0         &  0 &    \\
d_{\overline{\yng(2)}}   & 2\times(1,1,1,\overline{3})     &  0 &  0 &  0  & 0         &  0 &    \\
d_{\overline{\yng(1,1)}_{}} & 22\times(1,1,1,\overline{1}) &  0 &  0 &  0  & 0         &  0 &    \\
\hline
\end{array}$
\end{tabular}\end{table}

\begin{table}[htb]
\centering \footnotesize\renewcommand{\arraystretch}{1.3}
\caption{The spectrum of chiral and vector-like superfields, and their quantum numbers under the gauge symmetry ${\rm SU}(4)\times {\rm SU}(2)_L \times {\rm USp}(2)_R \times {\rm SU}(2)$ for the Model~\hyperref[model0]{1}.\\}\label{tab:spec0dual}
\begin{tabular}{c}$\begin{array}{|c|c||r|r|r||c|c|c|}\hline
\text{Model~\hyperref[model0dual]{1-dual}} & \text{Quantum Number}  & Q_4 & Q_{2L} & Q_{2R} & Q_{em} & B-L & \text{Field} \\
\hline\hline
ab               & 3 \times (\overline{4},2,1,1)           & -1 &  1 &  0  & \frac{1}{3}, -\frac{2}{3}, 1, 0 & -\frac{1}{3}, 1  &  Q_L, L_L\\
ac               & 3\times (4,1,\overline{2},1)            &  1 &  0 & -1  & -\frac{1}{3}, \frac{2}{3}, -1, 0  & \frac{1}{3}, -1  &  Q_R, L_R\\
bc               & 4\times (1,2,\overline{2},1)            &  0 &  1 & -1  & 1, 0, 0, -1 &  0  &  H^i_u, H^i_d\\
ad               & 4\times (4,1,1,\overline{2})            &  1 &  0 &  0  & \frac{1}{6}, -\frac{1}{2} & \frac{1}{3}, -1 &    \\
ad'              & 4\times (4,1,1,2)                       &  1 &  0 &  0  & \frac{1}{6}, -\frac{1}{2} & \frac{1}{3}, -1 &    \\
bd               & 4\times (1,2,1,\overline{2})            &  0 &  1 &  0  & \pm \frac{1}{2} & 0  &    \\
bd'              & 10\times (1,\overline{2},1,\overline{2}) &  0 & -1 &  0  & \mp \frac{1}{2} & 0  &    \\
cd               & 2\times (1,1,2,\overline{2})            &  0 &  0 &  1  & \pm \frac{1}{2} & 0  &    \\
cd'              & 2\times (1,1,\overline{2},\overline{2})&  0 &  0 & -1  & \mp \frac{1}{2} & 0  &    \\
a_{\overline{\yng(2)}}   & 2\times(\overline{10},1,1,1)    & -2 &  0 &  0  & \frac{1}{3}, -\frac{1}{3}, -1  & \frac{2}{3}, -2 &    \\
a_{\yng(1,1)}            & 2\times(6,1,1,1)                &  2 &  0 &  0  & \frac{1}{3}, -1  & \frac{2}{3}, -2 &    \\
b_{\yng(2)}              & 1\times(1,3,1,1)                &  0 &  2 &  0  & 0, \pm 1  &  0 &    \\
b_{\overline{\yng(1,1)}} & 1\times(1,\overline{1},1,1)     &  0 & -2 &  0  & 0         &  0 &    \\
d_{\overline{\yng(2)}}   & 2\times(1,1,1,\overline{3})     &  0 &  0 &  0  & 0         &  0 &    \\
d_{\overline{\yng(1,1)}_{}} & 22\times(1,1,1,\overline{1}) &  0 &  0 &  0  & 0         &  0 &    \\
\hline
\end{array}$
\end{tabular}\end{table}

Model~\hyperref[model0]{1} is a rare case without any filler branes and the hidden sector group SU(2) only comes from the $d$-stack of D6-branes. Its detailed particle spectrum is shown in table~\ref{tab:spec0}. There arise 4 Higgs particles from the $bc$ sector.

In addition to SM particles, we find several exotic particles charged under the hidden sector gauge group. However, in analogy with the quark condensation in QCD, the strong coupling dynamics of the hidden sector at some intermediate scale can provide a confining mechanism without affecting the anomaly matching conditions \cite{Cvetic:2002qa}. The resulting confined particle spectrum will be charged under the SM gauge group. Depending upon the type of the hidden sector gauge group, the particles can either self-confine to form composites of the meson-type or can mix with other gauge group factor to form particles of the baryon-type\footnote{Recall that in QCD, a meson is an inner product of a fundamental representation and an anti-fundamental representation while a baryon is a rank $3$ anti-symmetric representation of ${\rm SU}(3)_C$.}. Group theoretically, the meson-type composite is the pseudo inner product of two fundamental representations while the baryon-type composite is the rank 2N anti-symmetric representation for ${\rm USp}(2N)$ group with $N \geq 2$. Both kinds of these bound states transform as singlets under ${\rm USp}(2N)$.

By swapping the stacks $b$ and $c$ we can also obtain the dual models with the gauge group ${\rm SU}(4)_C \times {\rm SU}(2)_L \times {\rm USp}(2)_R $ where the left and right sectors are exchanged. This results in a completely different model after brane splitting as can be seen from the spectrum of the particles shown in table~\ref{tab:spec0dual} for the case of Model~\hyperref[model0dual]{1-dual}.

\begin{table}[htb]
\centering \footnotesize\renewcommand{\arraystretch}{1.3}
\caption{The chiral and vector-like superfields, and their quantum numbers under the gauge symmetry ${\rm SU}(4)\times {\rm USp}(2)_L \times {\rm SU}(2)_R \times {\rm SU}(2) \times {\rm USp}(2)$ for the Model~\hyperref[model2.1]{2}.\\} \label{tab:spec2.1}
\begin{tabular}{c}$\begin{array}{|c|c||r|r|r||c|c|c|}\hline
\rm{Model~\hyperref[model2.1]{2}} & \text{Quantum Number}  & Q_4 & Q_{2L} & Q_{2R} & Q_{em} & B-L & \text{Field} \\
\hline\hline
ab               & 3\times (4,\overline{2},1,1,1)            &  1 & -1 &  0 & -\frac{1}{3}, \frac{2}{3}, -1, 0 & \frac{1}{3}, -1  &  Q_L, L_L\\
ac'              & 3\times (\overline{4},1,\overline{2},1,1) & -1 &  0 & -1 & \frac{1}{3}, -\frac{2}{3}, 1, 0  & -\frac{1}{3}, 1  &  Q_R, L_R\\
bc               & 1 \times (1,2,\overline{2},1,1)           &  0 &  1 & -1 & 1, 0, 0, -1 &  0  &  H_u, H_d\\
bc'              & 1\times (1,\overline{2},\overline{2},1,1) &  0 & -1 & -1 & -1, 0, 0, 1 &  0  &  H'\\
ad               & 4\times (4,1,1,\overline{2},1)            &  1 &  0 &  0 & \frac{1}{6}, -\frac{1}{2} & \frac{1}{3}, -1 &    \\
ad'              & 4\times (4,1,1,2,1)                       &  1 &  0 &  0 & \frac{1}{6}, -\frac{1}{2} & \frac{1}{3}, -1 &    \\
bd               & 1\times (1,\overline{2},1,2,1)            &  0 & -1 &  0 & \mp \frac{1}{2} & 0  &    \\
bd'              & 1\times (1,2,1,2,1)                       &  0 &  1 &  0 & \pm \frac{1}{2} & 0  &    \\
cd               & 4\times (1,1,2,\overline{2},1)            &  0 &  0 &  1 & \pm \frac{1}{2} & 0  &    \\
cd'              & 4\times (1,1,\overline{2},\overline{2},1) &  0 &  0 & -1 & \mp \frac{1}{2} & 0  &    \\
c1               & 3\times (1,1,\overline{2},1,2)            &  0 &  0 & -1 & \mp \frac{1}{2} & 0  &    \\
d1               & 3\times (1,1,1,\overline{2},2)            &  0 &  0 &  0 &  0  &  0 &    \\
a_{\overline{\yng(2)}}   & 2\times(\overline{10},1,1,1,1)    & -2 &  0 &  0 &  \frac{1}{3}, -\frac{1}{3}, -1  & \frac{2}{3}, -2 &    \\
a_{\yng(1,1)}            & 2\times(6,1,1,1,1)                &  2 &  0 &  0 &  \frac{1}{3}, -1  & \frac{2}{3}, -2 &    \\
c_{\yng(2)}              & 2\times(1,1,3,1,1)                &  0 &  0 &  2 &  0, \pm 1  &  0 &    \\
c_{\overline{\yng(1,1)}} & 2\times(1,1,\overline{1},1,1)     &  0 &  0 & -2 &  0  &  0 &    \\
d_{\overline{\yng(1,1)}_{}} & 12\times(1,1,1,\overline{1},1) &  0 &  0 &  0 &  0  &  0 &    \\
\hline
\end{array}$
\end{tabular}\end{table}

Model~\hyperref[model2.1]{2} has the hidden sector group ${\rm SU(2)}\times {\rm USp(2)}$ coming from the $d$-stack of D6-branes and the $1$-stack of O6-planes. Its detailed particle spectrum is shown in table~\ref{tab:spec2.1}. There is single pair of Higgs from the $bc$ sector.

\begin{table}[htb]
\centering \footnotesize\renewcommand{\arraystretch}{1.3}
\caption{The chiral and vector-like superfields, and their quantum numbers under the gauge symmetry ${\rm SU}(4)\times {\rm USp}(2)_L \times {\rm SU}(2)_R \times {\rm SU}(2) \times {\rm USp}(2)$ for the Model~\hyperref[model2.2]{3}.\\} \label{tab:spec2.2}
\begin{tabular}{c}$\begin{array}{|c|c||r|r|r||c|c|c|}\hline
\rm{Model~\hyperref[model2.2]{3}} & \text{Quantum Number}  & Q_4 & Q_{2L} & Q_{2R} & Q_{em} & B-L & \text{Field} \\
\hline\hline
ab               & 3 \times (4,\overline{2},1,1,1)           &  1 & -1 &  0 & -\frac{1}{3}, \frac{2}{3}, -1, 0 & \frac{1}{3}, -1  &  Q_L, L_L\\
ac               & 3\times (\overline{4},1,2,1,1)            & -1 &  0 &  1 & \frac{1}{3}, -\frac{2}{3}, 1, 0  & -\frac{1}{3}, 1  &  Q_R, L_R\\
bc               & 1 \times (1,\overline{2},2,1,1)           &  0 & -1 &  1 & 1, 0, 0, -1 &  0  &  H_u, H_d\\
bc'              & 1\times (1,\overline{2},\overline{2},1,1) &  0 & -1 & -1 & -1, 0, 0, 1 &  0  &  H'\\
bd               & 6\times (1,2,1,\overline{2},1)            &  0 & -1 &  0 & \mp \frac{1}{2} & 0  &    \\
bd'              & 6\times (1,\overline{2},1,\overline{2},1) &  0 & -1 &  0 & \mp \frac{1}{2} & 0  &    \\
cd               & 3\times (1,1,2,\overline{2},1)            &  0 &  0 &  1 & \pm \frac{1}{2} & 0  &    \\
d2               & 1\times (1,1,1,2,\overline{2})            &  0 &  0 &  0 &  0  &  0 &    \\
c_{\overline{\yng(2)}}   & 2\times(1,1,\overline{3},1,1)     &  0 &  0 & -2 &  0, \pm 1  &  0 &    \\
c_{\yng(1,1)}            & 2\times(1,1,1,1,1)                &  0 &  0 &  2 &  0  &  0 &    \\
d_{\overline{\yng(2)}}   & 16\times(1,1,1,\overline{1},1)    &  0 &  0 &  0 &  0  &  0 &    \\
d_{\overline{\yng(1,1)}_{}} & 56\times(1,1,1,\overline{1},1) &  0 &  0 &  0 &  0  &  0 &    \\
\hline
\end{array}$
\end{tabular}\end{table}
\begin{table}[htb]
\centering \footnotesize
\caption{The composite particle spectrum for Model~\hyperref[model2.2]{3} formed due to the strong forces in hidden sector.\\} \label{exotic2.2}
\begin{tabular}{c}$\begin{array}{|c|c||c|c|}\hline
\multicolumn{2}{|c||}{\text{Model~\hyperref[model2.2]{3}}} & \multicolumn{2}{c|}{{\rm SU}(4)\times {\rm USp}(2)_L \times {\rm SU}(2)_R \times {\rm SU}(2) \times {\rm USp}(2)} \\
\hline \text{Confining Force} & \text{Intersection} & \text{Exotic Particle} & \text{Confined Particle Spectrum} \\
\hline\hline
{\rm USp}(2) & bd  & 6\times (1,2,1,\overline{2},1)            & 6\times(1,2^2,1,1,1),\, 18\times(1,\overline{2},2,1,1),\, \\
            & bd' & 6\times (1,\overline{2},1,\overline{2},1) & 6\times(1,\overline{2}^2,1,1,1),\, 18\times(1,2,2,1,1),\, \\
            & cd  & 3\times (1,1,2,\overline{2},1)            & 3\times(1,1,2^2,1,1) \\
\hline
\end{array}$
\end{tabular}\end{table}

Model~\hyperref[model2.2]{3} has the hidden sector group ${\rm SU(2)}\times {\rm USp(2)}$ coming from the $d$-stack of D6-branes and the $2$-stack of O6-planes. Its detailed particle spectrum is shown in table~\ref{tab:spec2.2}. Again, there is only a single pair of Higgs from the $bc$ sector. Various exotic particles can be decoupled either via self-confinement or mixed-confinement as shown in table~\ref{exotic2.2}.

\begin{table}[htb]
\centering \footnotesize\renewcommand{\arraystretch}{1.3}
\caption{The chiral and vector-like superfields, and their quantum numbers under the gauge symmetry ${\rm SU}(4)\times {\rm SU}(2)_L \times {\rm USp}(2)_R \times {\rm SU}(2) \times {\rm USp}(2)$ for the Model~\hyperref[model2.2dual]{3-dual}.\\} \label{tab:spec2.2dual}
\begin{tabular}{c}$\begin{array}{|c|c||r|r|r||c|c|c|}\hline
\text{Model~\hyperref[model2.2dual]{3-dual}} & \text{Quantum Number}  & Q_4 & Q_{2L} & Q_{2R} & Q_{em} & B-L & \text{Field} \\
\hline\hline
ab               & 3 \times (4,\overline{2},1,1,1)           &  1 & -1 &  0 & -\frac{1}{3}, \frac{2}{3}, -1, 0 & \frac{1}{3}, -1  &  Q_L, L_L\\
ac               & 3\times (\overline{4},1,2,1,1)            & -1 &  0 &  1 & \frac{1}{3}, -\frac{2}{3}, 1, 0  & -\frac{1}{3}, 1  &  Q_R, L_R\\
bc               & 1 \times (1,2,\overline{2},1,1)           &  0 &  1 & -1 & -1, 0, 0, 1 &  0  &  H_u, H_d\\
bd               & 3\times (1,2,1,\overline{2},1)            &  0 &  1 &  0 & \pm \frac{1}{2} & 0  &    \\
cd               & 6\times (1,1,2,\overline{2},1)            &  0 &  0 & -1 & \mp \frac{1}{2} & 0  &    \\
cd'              & 6\times (1,1,\overline{2},\overline{2},1) &  0 &  0 & -1 & \mp \frac{1}{2} & 0  &    \\
d2               & 1\times (1,1,1,2,\overline{2})            &  0 &  0 &  0 &  0  &  0 &    \\
b_{\overline{\yng(2)}}   & 2\times(1,\overline{3},1,1,1)     &  0 & -2 &  0 &  0, \pm 1  &  0 &    \\
b_{\yng(1,1)}            & 2\times(1,1,1,1,1)                &  0 &  2 &  0 &  0  &  0 &    \\
d_{\overline{\yng(2)}}   & 16\times(1,1,1,\overline{1},1)    &  0 &  0 &  0 &  0  &  0 &    \\
d_{\overline{\yng(1,1)}_{}} & 56\times(1,1,1,\overline{1},1) &  0 &  0 &  0 &  0  &  0 &    \\
\hline
\end{array}$
\end{tabular}\end{table}
\begin{table}[htb]
\centering \footnotesize
\caption{The composite particle spectrum for Model~\hyperref[model2.2dual]{3-dual} formed due to the strong forces in hidden sector.\\} \label{exotic2.2dual}
\begin{tabular}{c}$\begin{array}{|c|c||c|c|}\hline
\multicolumn{2}{|c||}{\text{Model~\hyperref[model2.2dual]{3-dual}}} & \multicolumn{2}{c|}{{\rm SU}(4)\times {\rm SU}(2)_L \times {\rm USp}(2)_R \times {\rm SU}(2) \times {\rm USp}(2)} \\
\hline \text{Confining Force} & \text{Intersection} & \text{Exotic Particle} & \text{Confined Particle Spectrum} \\
\hline\hline
{\rm USp}(2) & bd  & 3\times (1,2,1,\overline{2},1)            & 3\times(1,2^2,1,1,1) \\
             & cd  & 6\times (1,1,2,\overline{2},1)            & 6\times(1,1,2^2,1,1),\, 18\times(1,2,2,1,1),\, \\
             & cd' & 6\times (1,1,\overline{2},\overline{2},1) & 6\times(1,1,\overline{2}^2,1,1),\, 18\times(1,2,\overline{2},1,1),\, \\
\hline
\end{array}$
\end{tabular}\end{table}

By swapping the stacks $b$ and $c$ we obtain the Model~\hyperref[model2.2dual]{3-dual} with the gauge group ${\rm SU}(4)_C \times {\rm SU}(2)_L \times {\rm USp}(2)_R $. The hidden sector group ${\rm SU(2)}\times {\rm USp(2)}$ comes from the $d$-stack of D6-branes and the $2$-stack of O6-planes. Its detailed particle spectrum is shown in table~\ref{tab:spec2.2dual}. Again, there is only a single pair of Higgs from the $bc$ sector. Various exotic particles can be decoupled either via self-confinement or mixed-confinement as shown in table~\ref{exotic2.2dual}.

\begin{table}[htb]
\centering \footnotesize\renewcommand{\arraystretch}{1.3}
\caption{The chiral and vector-like superfields, and their quantum numbers under the gauge symmetry ${\rm SU}(4)\times {\rm USp}(2)_L \times {\rm SU}(2)_R \times {\rm SU}(2) \times {\rm USp}(4)$ for the Model~\hyperref[model4]{4}.\\}\label{tab:spec4}
\begin{tabular}{c}$\begin{array}{|c|c||r|r|r||c|c|c|}\hline
\rm{Model~\hyperref[model4]{4}} & \text{Quantum Number}      & Q_4 & Q_{2L} & Q_{2R} & Q_{em} & B-L & \text{Field} \\
\hline\hline
ab               & 3 \times (4,\overline{2},1,1,1)           &  1  & -1  &  0  & -\frac{1}{3}, \frac{2}{3}, -1, 0 & \frac{1}{3}, -1  &  Q_L, L_L\\
ac               & 3\times (\overline{4},1,2,1,1)            & -1  &  0  &  1  & \frac{1}{3}, -\frac{2}{3}, 1, 0  & -\frac{1}{3}, 1  &  Q_R, L_R\\
bc               & 2\times (1,\overline{2},2,1,1)            &  0  & -1  &  1  &  1, 0, 0, -1 &  0  &  H^i_u, H^i_d\\
bc'              & 2\times (1,\overline{2},\overline{2},1,1) &  0  & -1  & -1  &  -1, 0, 0, 1 &  0  &  H'^i\\
ad               & 4\times (4,1,1,\overline{2},1)            &  1  &  0  &  0  &  \frac{1}{6}, -\frac{1}{2} & \frac{1}{3}, -1 &    \\
ad'              & 4\times (4,1,1,2,1)                       &  1  &  0  &  0  &  \frac{1}{6}, -\frac{1}{2} & \frac{1}{3}, -1 &    \\
bd               & 2\times (1,2,1,\overline{2},1)            &  0  &  1  &  0  &  \pm \frac{1}{2} & 0  &    \\
bd'              & 2\times (1,\overline{2},1,\overline{2},1) &  0  & -1  &  0  &  \mp \frac{1}{2} & 0  &    \\
cd               & 8\times (1,1,2,\overline{2},1)            &  0  &  0  &  1  &  \pm \frac{1}{2} & 0  &    \\
cd'              & 8\times (1,1,\overline{2},\overline{2},1) &  0  &  0  & -1  &  \mp \frac{1}{2} & 0  &    \\
d2               & 1\times (1,1,1,2,\overline{4})            &  0  &  0  &  0  &  0  &  0 &    \\
a_{\overline{\yng(2)}}   & 2\times(\overline{10},1,1,1,1)    & -2 & 0 &  0  &  \frac{1}{3}, -\frac{1}{3}, -1  & \frac{2}{3}, -2 &    \\
a_{\yng(1,1)}            & 2\times(6,1,1,1,1)                &  2 & 0 &  0  &  \frac{1}{3}, -1  & \frac{2}{3}, -2 &    \\
c_{\overline{\yng(2)}}   & 1\times(1,1,\overline{3},1,1)     &  0 & 0 & -2  &  0, \pm 1  &  0 &    \\
c_{\yng(1,1)}            & 1\times(1,1,\overline{1},1,1)     &  0 & 0 &  2  &  0  &  0 &    \\
d_{\overline{\yng(2)}}   & 2\times(1,1,1,\overline{3},1)     &  0 & 0 &  0  &  0  &  0 &    \\
d_{\overline{\yng(1,1)}_{}} & 22\times(1,1,1,\overline{1},1) &  0 & 0 &  0  &  0  &  0 &    \\
\hline
\end{array}$
\end{tabular}\end{table}

Model~\hyperref[model4]{4} has the hidden sector group ${\rm SU(2)}\times {\rm USp(4)}$ coming from the $d$-stack of D6-branes and the $2$-stack of O6-planes. Its detailed particle spectrum is shown in table~\ref{tab:spec4}. There arise two pairs of Higgs from the $bc$ sector.

\begin{table}[htb]
\centering \footnotesize\renewcommand{\arraystretch}{1.3}
\caption{The chiral and vector-like superfields, and their quantum numbers under the gauge symmetry ${\rm SU}(4)\times {\rm USp}(2)_L \times {\rm SU}(2)_R \times {\rm SU}(2) \times {\rm USp}(6)$ for the Model~\hyperref[model6]{5}.\\}\label{tab:spec6}
\begin{tabular}{c}$\begin{array}{|c|c||r|r|r||c|c|c|}\hline
\rm{Model~\hyperref[model6]{5}} & \text{Quantum Number}      & Q_4 & Q_{2L} & Q_{2R} & Q_{em} & B-L & \text{Field} \\
\hline\hline
ab               & 3 \times (4,\overline{2},1,1,1)           &  1  & -1  &  0  & -\frac{1}{3}, \frac{2}{3}, -1, 0 & \frac{1}{3}, -1  &  Q_L, L_L\\
ac               & 3\times (\overline{4},1,2,1,1)            & -1  &  0  &  1  & \frac{1}{3}, -\frac{2}{3}, 1, 0  & -\frac{1}{3}, 1  &  Q_R, L_R\\
bc               & 1\times (1,2,\overline{2},1,1)            &  0  &  1  & -1  &  1, 0, 0, -1 &  0  &  H_u, H_d\\
bc'              & 1\times (1,2,2,1,1)                       &  0  & -1  & -1  &  -1, 0, 0, 1 &  0  &  H'\\
ad               & 4\times (4,1,1,\overline{2},1)            &  1  &  0  &  0  &  \frac{1}{6}, -\frac{1}{2} & \frac{1}{3}, -1 &    \\
ad'              & 4\times (4,1,1,2,1)                       &  1  &  0  &  0  &  \frac{1}{6}, -\frac{1}{2} & \frac{1}{3}, -1 &    \\
bd               & 2\times (1,2,1,\overline{2},1)            &  0  &  1  &  0  &  \pm \frac{1}{2} & 0  &    \\
bd'              & 2\times (1,\overline{2},1,\overline{2},1) &  0  & -1  &  0  &  \mp \frac{1}{2} & 0  &    \\
cd               & 7\times (1,1,2,\overline{2},1)            &  0  &  0  &  1  &  \pm \frac{1}{2} & 0  &    \\
cd'              & 10\times (1,1,\overline{2},\overline{2},1)&  0  &  0  &  1  &  \pm \frac{1}{2} & 0  &    \\
d3               & 1\times (1,1,1,\overline{2},6)            &  0  &  0  &  0  &  0  &  0 &    \\
a_{\overline{\yng(2)}}   & 2\times(\overline{10},1,1,1,1)    & -2 & 0 &  0  &  \frac{1}{3}, -\frac{1}{3}, -1  & \frac{2}{3}, -2 &    \\
a_{\yng(1,1)}            & 2\times(6,1,1,1,1)                &  2 & 0 &  0  &  \frac{1}{3}, -1  & \frac{2}{3}, -2 &    \\
c_{\yng(2)}              & 1\times(1,1,3,1,1)                &  0 & 0 &  2  &  0, \pm 1  &  0 &    \\
c_{\yng(1,1)}            & 1\times(1,1,\overline{1},1,1)     &  0 & 0 &  2  &  0  &  0 &    \\
d_{\overline{\yng(2)}}   & 2\times(1,1,1,\overline{3},1)     &  0 & 0 &  0  &  0  &  0 &    \\
d_{\overline{\yng(1,1)}_{}} & 22\times(1,1,1,\overline{1},1) &  0 & 0 &  0  &  0  &  0 &    \\
\hline
\end{array}$
\end{tabular}\end{table}

Model~\hyperref[model6]{5} has the hidden sector group ${\rm SU(2)}\times {\rm USp(6)}$ coming from the $d$-stack of D6-branes and the $3$-stack of O6-planes. Its detailed particle spectrum is shown in table~\ref{tab:spec6}. There only arises a single pair of Higgs from the $bc$ sector.

In general, there exist exotic particles in the string model building, and thus we need to discuss how to decouple these exotics from the low energy spectrum. For the Models \hyperref[model2.2]{3} and \hyperref[model2.2dual]{3-dual}, similar to refs.~\cite{Chen:2007px,Chen:2007zu}, we can decouple all the exotic particles. However, because there exist the chiral multiplets under ${\rm SU}(4)_C$ symmetric representation in the rest of the models, it seems to us that we might not decouple all the exotic particles in these models, {\it i.e.}, Models \hyperref[model0]{1}, \hyperref[model2.1]{2}, \hyperref[model4]{4}, \hyperref[model6]{5} and their duals. In the subsequent section~\ref{sec:g-c-u}, we will study how to realize the string-scale gauge coupling relations in Models 1 and 3, and assume that all the exotic particles in Model 1 can be decoupled for simplicity.

\FloatBarrier
\section{Gauge coupling relations}\label{sec:g-c-u}
Let us point out an interesting subtlety under the exchange of left and right sectors by swapping the stacks $b$ and $c$ to yield the gauge group ${\rm SU}(4)_C \times {\rm SU}(2)_L \times {\rm USp}(2)_R $. Although this results in an equivalent Pati-Salam model with similar\footnote{The low energy spectrum is obviously different since standard model is chiral.} high energy particle spectra, however the gauge coupling relation does not remain invariant unless the model exhibits exact gauge coupling unification. None of the models in our search exhibits exact gauge coupling relation as shown in the captions of the tables listed in the appendix \ref{appendix}. For instance in Model~\hyperref[model0]{1} the gauge coupling relation under the exchange of left and the right sectors will be transformed as,
\begin{align}
{\rm Model}~\hyperref[model0]{1} \qquad g^2_a &=\frac{8}{7}\, g^2_b=\frac{4}{3}\, g_c^2= \frac{6}{5}\,\left(\frac{5}{3}\,g^2_Y\right)= \frac{32}{7}\sqrt[4]{\frac{2}{3}}\, \pi \,e^{\phi_4} \nonumber\\
\xrightarrow{~b \leftrightarrow c~} g^2_a &=\frac{4}{3}\, g^2_b=\frac{8}{7}\, g_c^2= \frac{38}{35}\,\left(\frac{5}{3}\,g^2_Y\right)= \frac{32}{7}\sqrt[4]{\frac{2}{3}}\, \pi \,e^{\phi_4}.
\end{align}
Therefore, in models with approximate gauge coupling unification this asymmetry under the exchange of left and right sectors may be exploited to refine the gauge coupling relations as exemplified in \cite{He:2021kbj}. Similarly, the gauge coupling relations for other models are given as follows:

\begin{align}
{\rm Model}~\hyperref[model2.1]{2} \qquad g^2_a &=\frac{11}{36}\, g^2_b=\frac{1}{3}\, g_c^2= \frac{3}{5}\,\left(\frac{5}{3}\,g^2_Y\right)= \frac{2 \sqrt{2}}{9} 11^{3/4}\, \pi \,e^{\phi_4} \nonumber\\
\xrightarrow{~b \leftrightarrow c~} g^2_a &=\frac{1}{3}\, g^2_b=\frac{11}{36}\, g_c^2= \frac{7}{12}\,\left(\frac{5}{3}\,g^2_Y\right)= \frac{2 \sqrt{2}}{9} 11^{3/4}\, \pi \,e^{\phi_4},
\end{align}

\begin{align}
{\rm Model}~\hyperref[model2.2]{3} \qquad g^2_a &=\frac{5}{18}\, g^2_b=\frac{1}{3}\, g_c^2= \frac{3}{5}\,\left(\frac{5}{3}\,g^2_Y\right)= \frac{4 \sqrt{2}}{9} 5^{3/4} \, \pi \,e^{\phi_4} \nonumber\\
\xrightarrow{~b \leftrightarrow c~} g^2_a &=\frac{1}{3}\, g^2_b=\frac{5}{18}\, g_c^2= \frac{17}{30}\,\left(\frac{5}{3}\,g^2_Y\right)= \frac{4 \sqrt{2}}{9} 5^{3/4} \, \pi \,e^{\phi_4},
\end{align}

\begin{align}
{\rm Model}~\hyperref[model4]{4} \qquad g^2_a &=\frac{11}{18}\, g^2_b=\frac{2}{3}\, g_c^2= \frac{4}{5}\,\left(\frac{5}{3}\,g^2_Y\right)= \frac{4}{9}\, 11^{3/4}\, \pi \,e^{\phi_4} \nonumber\\
\xrightarrow{~b \leftrightarrow c~} g^2_a &=\frac{2}{3}\, g^2_b=\frac{11}{18}\, g_c^2= \frac{23}{30}\,\left(\frac{5}{3}\,g^2_Y\right)= \frac{4}{9}\, 11^{3/4}\, \pi \,e^{\phi_4},
\end{align}

\begin{align}
{\rm Model}~\hyperref[model6]{5} \qquad g^2_a &=\frac{7}{22}\, g^2_b=\frac{1}{3}\, g_c^2= \frac{3}{5}\,\left(\frac{5}{3}\,g^2_Y\right)= \frac{4\sqrt{2}}{11}\frac{7^{3/4}}{\sqrt[4]{3}}\, \pi \,e^{\phi_4} \nonumber\\
\xrightarrow{~b \leftrightarrow c~} g^2_a &=\frac{1}{3}\, g^2_b=\frac{7}{22}\, g_c^2= \frac{13}{22}\,\left(\frac{5}{3}\,g^2_Y\right)= \frac{4\sqrt{2}}{11}\frac{7^{3/4}}{\sqrt[4]{3}}\, \pi \,e^{\phi_4}.
\end{align}


Gauge coupling unification has been extensively studied in refs.~\cite{Bachas:1995yt, Lopez:1996gd, Blumenhagen:2003jy, Barger:2005qy, Jiang:2006hf, Barger:2007qb, Jiang:2008xrg, Jiang:2009za, Kokorelis:2016ckp, Chen:2017rpn, Chen:2018ucf, He:2021gug}. In the minimal supersymmetric standard model (MSSM), the gauge couplings unify at the GUT scale, $2\times 10^{16}$ GeV which is one order smaller than string scale $M_{\rm string} \simeq  5\times 10^{17}~{\rm GeV}$~\cite{Ellis:1990wk, Langacker:1991an, Amaldi:1991cn}. Realizing string-scale gauge coupling thus becomes an important question in string phenomenology.

Among the five models and their duals under the $b \leftrightarrow c$ exchange, Model~\hyperref[model0]{1}, Model~\hyperref[model0dual]{1-dual}, Model~\hyperref[model2.2]{3} and Model~\hyperref[model2.2]{3-dual} are particularly interesting. While the Models \hyperref[model0]{1} and \hyperref[model0dual]{1-dual} contain the largest wrapping number of 4, the exotic particles in Models~\hyperref[model2.2]{3} and ~\hyperref[model2.2]{3-dual} can be decoupled via Higgs mechanism and instanton effects \cite{Blumenhagen:2006xt, Haack:2006cy, Florea:2006si, Chen:2007px, Chen:2007zu} as discussed earlier. Henceforth, we choose these four models \textit{viz.} Model~\hyperref[model0]{1}, Model~\hyperref[model0dual]{1-dual}, Model~\hyperref[model2.2]{3} and Model~\hyperref[model2.2]{3-dual} to illustrate the string-scale gauge coupling relations in the supersymmetric Pati-Salam models utilizing two-loop renormalization group equations (RGEs)~\cite{Barger:2005qy, Barger:2007qb}.

In order to achieve the string-scale gauge coupling unification, we generally introduce new particles at the intermediate scale either from the adjoint representations of ${\rm SU}(4)_C$ and ${\rm SU}(2)_L$ gauge symmetries, SM vector-like particles from four-dimensional chiral sectors, or from vector-like particles from ${\cal N} = 2$ subsector. These extra particles naturally emerge from string model building.

\begin{figure}[htb]
    \centering
    \includegraphics[width=0.45\linewidth]{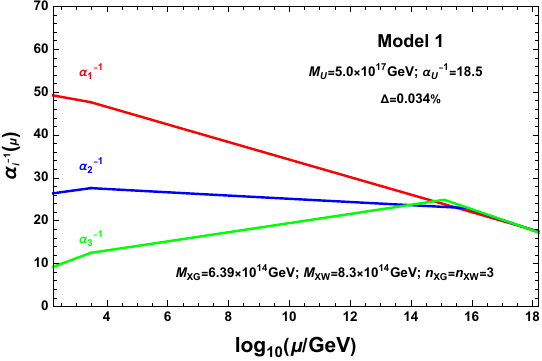}\\
    \includegraphics[width=0.45\linewidth]{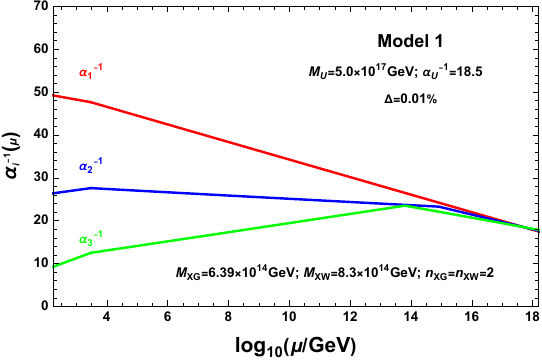}\qquad
    \includegraphics[width=0.45\linewidth]{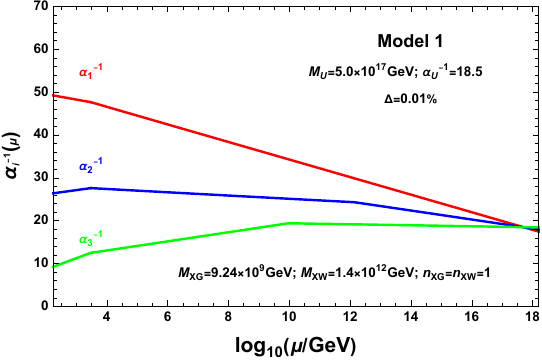}
    \caption{String-scale gauge coupling relations for Model~\hyperref[model0]{1} with $XG+XW$. The number of added particle pairs is (a) 3, (b) 2, and (c) 1.}
    \label{fig:model1}
\end{figure}
\begin{figure}[thb]
    \centering
    \includegraphics[width=0.45\linewidth]{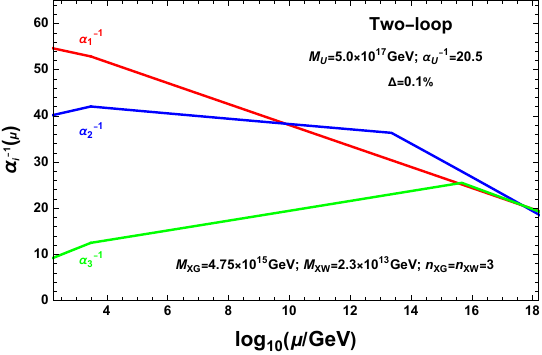}\\
    \includegraphics[width=0.45\linewidth]{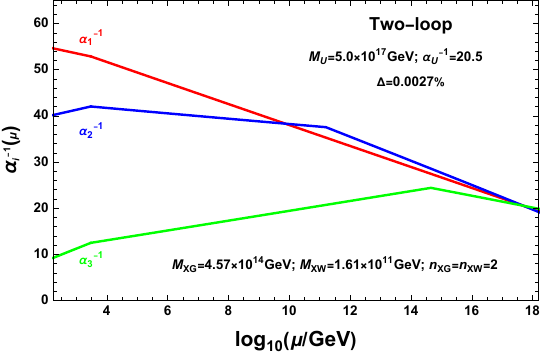}\qquad
    \includegraphics[width=0.45\linewidth]{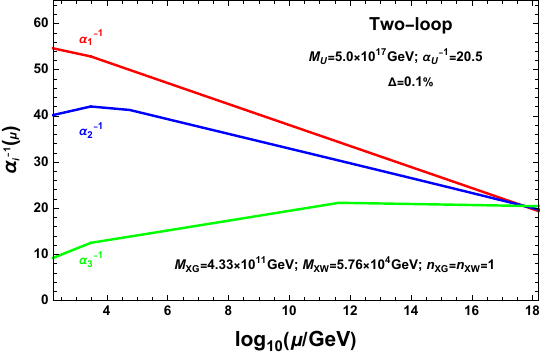}
    \caption{String-scale gauge coupling relations for Model~\hyperref[model0dual]{1-dual} with $XG+XW$. The number of added particle pairs is (a) 3, (b) 2, and (c) 1. }
    \label{fig:model1dual}
\end{figure}

Following the conventions in refs.~\cite{Chen:2017rpn, Chen:2018ucf, He:2021kbj, Li:2022cqk} and taking the supersymmetry breaking scale at $3$~TeV, the gauge couplings unify close to the string scale, $5\times10^{17}$~GeV.
It is customary to redefine the set of parameters as,
\begin{equation}
    \alpha_1\equiv k_Yg_Y^2/4\pi, \quad \alpha_2 \equiv k_2g_b^2/4\pi, \quad \alpha_3 \equiv g_a^2/4\pi.
\end{equation}
where $k_Y$ is canonically chosen to be $5/3$. In our calculations, the The string-scale gauge coupling relation in the evolution is realized via $\alpha_{\text{U}}^{-1}\equiv \alpha_1^{-1}=(\alpha_2^{-1}+\alpha_3^{-1})/2$ and the deviation $\Delta=|\alpha_1^{-1}-\alpha_2^{-1}|/\alpha_1^{-1}$ is limited to within 0.1\%.

\begin{figure}[thb]
    \centering
    \includegraphics[width=0.45\linewidth]{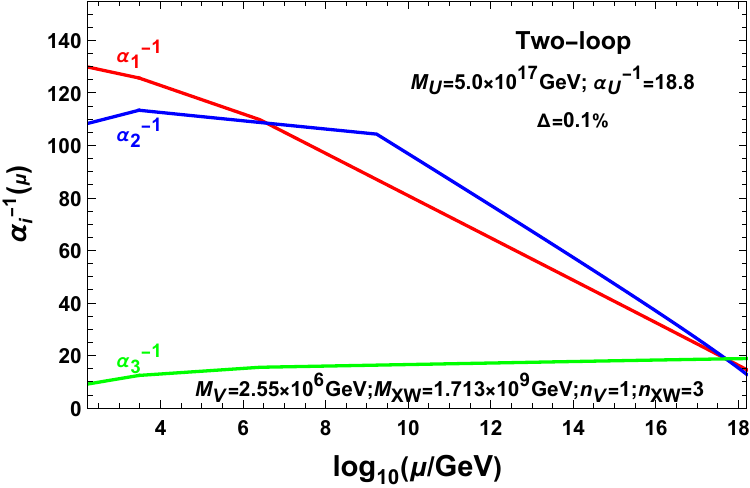}\qquad
    \includegraphics[width=0.45\linewidth]{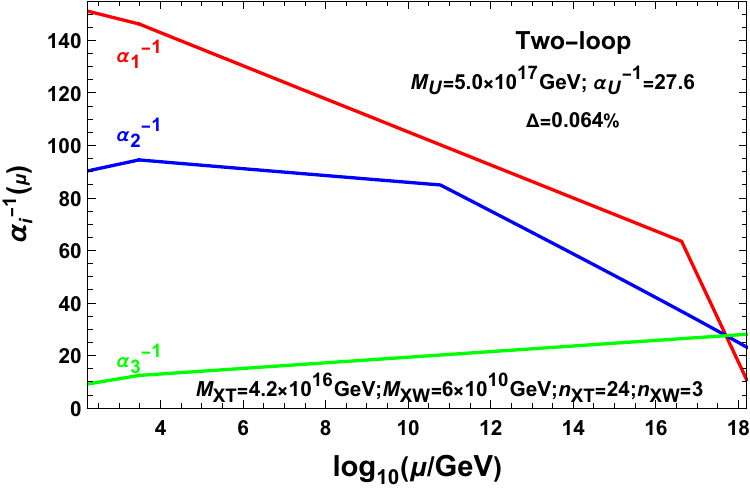}
    \caption{String-scale gauge coupling relations for Model~\hyperref[model2.2]{3} (a) and Model~\hyperref[model2.2dual]{3-dual} (b) with vector-like particles. (a) $V$ represents $XD+\overline{XD}$, $XU+\overline{XU}$, and $XE+\overline{XE}$; (b) $XT$ represents $XT+\overline{XT}$.}
    \label{fig:model3_3dual}
\end{figure}

For Model~\hyperref[model0]{1} and Model~\hyperref[model0dual]{1-dual} we use $XG$ and $XW$ particles coming from the chiral adjoints associated with the special unitary group factors in the gauge group having the quantum numbers, $XG=(\mathbf{8,1,0})$ and $XW=(\mathbf{1,3,0})$, respectively. The one-loop contributions from these two particles to the beta function are $\Delta b(XG)=(0,0,3)$ and $\Delta b(XW)=(0,2,0)$ respectively \cite{Jiang:2006hf, Barger:2007qb}. As the number of particles $n_{XV}(XV=XG,XW)$ increases, the mass of the particles $M_{XV}$ increases and the splitting between the $XW$ and $XG$ masses decreases.

In figure~\ref{fig:model1} and figure~\ref{fig:model1dual} we display the results of two-loop RGE running of the gauge couplings for Model~\hyperref[model0]{1} and Model~\hyperref[model0dual]{1-dual} by adding 3 pairs, 2 pairs and 1 pair of $XG$ and $XW$ particles respectively. Above $M_{XV}$, the running of ${\rm SU}(3)_C$ coupling and ${\rm SU}(2)_L$ coupling are reducing due to the non-zero beta functions $\Delta b_3(XG)$ and $\Delta b_2(XW)$.

Similarly in figure~\ref{fig:model3_3dual} we show the corresponding results of two-loop RGE running of the gauge couplings for Model~\hyperref[model2.2]{3} and Model~\hyperref[model2.2dual]{3-dual} by adding vector-like particles. For the Model \hyperref[model2.2]{3}, the exotic particles are $XD+\overline{XD}$, $XU+\overline{XU}$, $XE+\overline{XE}$ and $XW$ particles.
Under the $SU(3)_C\times SU(2)_L\times U(1)_Y$ gauge symmetry, the corresponding quantum numbers and the contributions to one-loop level beta functions are
\begin{alignat}{4}
    XD&=\mathbf{({3},1, -{1\over 3})}, &\quad \overline{XD}&=\mathbf{({\bar 3},  1, {1\over 3})},  &\qquad \Delta b &= 	({2\over 5}, 0, 1)\,;\nonumber\\
    XU&=\mathbf{({3},1, {2\over 3})},  &\quad \overline{XU}&=\mathbf{({\bar 3},  1, -{2\over 3})}, &\qquad  \Delta b &=	({8\over 5}, 0, 1)\,;\nonumber\\
    XE&=\mathbf{({1},1, {1})},         &\quad \overline{XE}&=\mathbf{({1},1,-{1})},                &\qquad  \Delta b &= ({6\over 5}, 0, 0)\,.
\end{alignat}
In the supersymmetric Pati-Salam models, these vector-like particles emerge from ${\cal N}=2$ subsector intersections between $a$ and $c'$ stacks of D6-brane. The number of these particles is thus $n_V=I_{ac'}$.
The $XT+\overline{XT}$ from $cd$ and $cd'$ sectors are added to Model \hyperref[model2.2]{3-dual} to achieve the string-scale gauge coupling relations. The quantum numbers of $XT+\overline{XT}$ under the $U(4)\times USp(2)_L\times U(2)_R\times U(2)$ gauge symmetry are $\mathbf{(1,1,2,2,1)+(1,1,2,\Bar{2},1)}$, and under the $SU(3)_C\times SU(2)_L\times U(1)_Y$ gauge symmetry are $\mathbf{(1,1,{1\over 2})+(1,1,-{1\over 2})}$. The contributions to one-loop level beta functions are $\Delta b=({6\over 5},0,0)$. Since $I_{cd}=I_{cd'}=6$, we have $6\times 2\times 2=24$ pairs of $XT+\overline{XT}$.
Therefore, we are able to show string-scale gauge coupling relations for each of the four chosen models.

\FloatBarrier
\section{Discussion and conclusion}\label{sec:conclusion}
We have found new three-family supersymmetric Pati-Salam models with the gauge group ${\rm SU}(4)_C \times {\rm USp}(2)_L \times {\rm SU}(2)_R$ on a $\mathbbm{T}^6/(\mathbbm{Z}_2\times \mathbbm{Z}_2)$ orientifold from intersecting D6-branes at angles from the perspective of from IIA string theory without fluxes for the first time. Unlike the special unitary group, the symplectic group is a simpler choice as there is no associated global anomalous U(1) group. The unitary symplectic group arises by placing the D6-brane stack parallel to any of the four O6-planes. This seemingly minor change from SU(2) to USp(2) group is quite restrictive and drastically reduces the number of consistent three-family models. Therefore, while the number of three-family models with ${\rm SU}(4)_C \times {\rm SU}(2)_L \times {\rm SU}(2)_R$ groups is found to be 202,752~\cite{He:2021kbj,He:2021gug}, the similar models by replacing either one of the SU(2) factors with a USp(2) is only 5. It is interesting to compare our search with ref.~\cite{Cvetic:2004nk} where one of us (TL) found only a single consistent three-family supersymmetric Pati-Salam model with the specific gauge group ${\rm SU}(4)_C \times {\rm USp}(2)_L \times {\rm USp}(2)_R$.

We have displayed the perturbative particle spectra of all the five inequivalent representative models and also discussed the duality under the exchange of the left and the right sectors. In particular, one of models (Model~\hyperref[model0]{1}) does not contain any filler O6-planes whereas the other four models contain a single stack of filler branes each. The highest wrapping number is found to be 4 and all models contain extra exotic particles. We have shown the decoupling of such exotic particles originating from the hidden sector through the mechanism of strong coupling dynamics for the Model~\hyperref[model2.2]{3} and Model~\hyperref[model2.2dual]{3-dual}. We have also calculated the tree-level gauge coupling relations for all models. By running the two-loop RGEs with the appropriate addition of ${\rm SU}(4)_C$/${\rm SU}(2)_L$ chiral adjoints and vector-like particle pairs, the gauge couplings are shown to be unified at the string-scale for some of the models.

Among the five models, the highest number of Higgs pairs is found to be 4 for Model~\hyperref[model0]{1} whereas all other models either contain a double or a single pair each. Thus, Model~\hyperref[model0]{1} may be the only viable candidate to explain the Yukawa couplings and fermion masses. However, we note that the two exotic symmetric representations corresponding to the ${\rm SU}(4)_C$ group in Model~\hyperref[model0]{1} cannot be decoupled from the visible spectrum. It is Model~\hyperref[model2.2]{3} and Model~\hyperref[model2.2dual]{3-dual} where all exotic states can be decoupled from the visible spectrum since there exist no exotic symmetric representations corresponding to the ${\rm SU}(4)_C$ group. However, both Model~\hyperref[model2.2]{3} and Model~\hyperref[model2.2dual]{3-dual} suffer from the rank-1 problem as the number of Higgs pair is only one each. In a related paper  \cite{Mansha:2023kwq}, we also discuss similar construction of supersymmetric Pati-Salam models with symplectic group from the perspective of possible hidden sector variations that can also affect the gauge coupling relations at the string-scale.

\acknowledgments{ TL is supported in part by the National Key Research and Development Program of China Grant No. 2020YFC2201504, by the Projects No. 11875062, No. 11947302, No. 12047503, and No. 12275333 supported by the National Natural Science Foundation of China, by the Key Research Program of the Chinese Academy of Sciences, Grant No. XDPB15, by the Scientific Instrument Developing Project of the Chinese Academy of Sciences, Grant No. YJKYYQ20190049, and by the International Partnership Program of Chinese Academy of Sciences for Grand Challenges, Grant No. 112311KYSB20210012. L Wu is supported in part by the Natural Science Basic Research Program of Shaanxi, Program No. 2024JC-YBMS-039. }

\appendix

\section{Three-family Pati-Salam models with a symplectic group}\label{appendix}

In this appendix, we list all representative three-family supersymmetric Pati-Salam models with a symplectic group obtained from random scanning method. $a, b, c, d$ in the first column in every table represent the four stacks of D6-branes, respectively. Similarly, $1, 2, 3, 4$ in the first columns is a short-handed notation for the filler branes along the $\Omega {\cal R}$, $\Omega {\cal R} \omega$, $\Omega {\cal R} \theta \omega$ and $\Omega {\cal R} \theta$ O6-planes, respectively. The second column in each table lists the numbers of D6-branes in the respective stack. In the third column we record the wrapping numbers of each D6-brane configuration. The rest of the columns record the intersection numbers between various stacks. For instance, in the $b$ column of table~\ref{model0}, from top to bottom, the numbers represent intersection numbers $I_{ab}, I_{bc}, I_{bd}$, etc.  As usual, $b'$ and $c'$ are the orientifold $\Omega {\cal R}$ image of $b$ and $c$ stacks of D6-branes. We also list the relation between $x_A, x_B, x_C, x_D$, which are determined by the supersymmetry conditions~\eqref{susyconditions}, as well as the relation between the moduli parameter $\chi_1,\, \chi_2,\, \chi_3$. The one loop beta functions $\beta^g_i$ for each of the hidden sector stack is also listed. The gauge coupling relations are given in the caption of each table.

\begin{table}[htb]  
\centering \footnotesize
\caption{D6-brane configurations and intersection numbers of Model~\hyperref[model0]{1}, and its gauge coupling relation is
$g^2_a=\frac{8}{7}\, g^2_b=\frac{4}{3}\, g_c^2= \frac{6}{5}\,\left(\frac{5}{3}\,g^2_Y\right)= \frac{32}{7}\sqrt[4]{\frac{2}{3}}\, \pi \,e^{\phi_4}$.\\} \label{model0}
\begin{tabular}{c}$\begin{array}{|c|c|c |r|r|r| r|r|r| r|r|r| r|r|r|} 	\hline 	\multicolumn{3}{|c|}{\text{Model~\hyperref[model0]{1}}}  & \multicolumn{11}{c|}{{\rm U}(4)\times {\rm USp}(2)_L \times {\rm U}(2)_R \times {\rm U}(2)} \\
	\hline \hline \rm{stack} & N & (n^1,l^1)\times (n^2,l^2)\times (n^3,l^3) & n_{\yng(2)} & n_{\yng(1,1)_{}} & b & c & c'& d & d' & 1 & 2 & 3 & 4 \\
	\hline
	a   & 8   & (-1, -3)\times (0, -1)\times (-1, -1) & -2    & 2     & 3    & -3    & 0     & 4     & 4     & 0     & 0     & 0    & 0 \\
    b   & 2   & (1, 0)\times (1, 0)\times (2, 0)      & 0     & 0     & \text{-} & 4     & -4    & 2     & -2    & 0     & 0     & 0    & 0 \\
    c   & 4   & (0, 1)\times (3, -4)\times (1, -1)    & 1     & -1    & \text{-} & \text{-}  & \text{-}  & 4     & -10   & 0     & 0     & 0    & 0 \\
    d   & 4   & (-1, 1)\times (1, -2)\times (-3, -1)  & -2    & -22   & \text{-} & \text{-}  & \text{-}  & \text{-}  & \text{-}  & 0     & 0     & 0    & 0 \\
	\hline
          &     &                     &  \multicolumn{11}{c|}{\beta^g_{d}=8, \quad  x_A=\frac{3}{4}x_B=\frac{1}{3}x_C=\frac{1}{24}x_D} \\
          &     &                     &  \multicolumn{11}{c|}{\chi_1=\sqrt{6}/18, \quad \chi_2=\sqrt{6}/8, \quad \chi_3= 2\sqrt{6}} \\
	\hline
\end{array}$
\end{tabular}\end{table}

\begin{table}[htb]  
\centering \footnotesize
\caption{D6-brane configurations and intersection numbers of Model~\hyperref[model0dual]{1-dual}, and its gauge coupling relation is
$g^2_a=\frac{4}{3}\, g^2_b=\frac{8}{7}\, g_c^2= \frac{38}{35}\,\left(\frac{5}{3}\,g^2_Y\right)= \frac{32}{7}\sqrt[4]{\frac{2}{3}}\, \pi \,e^{\phi_4}$.\\} \label{model0dual}
\begin{tabular}{c}$\begin{array}{|c|c|c |r|r|r| r|r|r| r|r|r| r|r|r|}
	\hline
	\multicolumn{3}{|c|}{\text{Model~\hyperref[model0dual]{1-dual}}} & \multicolumn{11}{c|}{{\rm U}(4)\times {\rm U}(2)_L \times {\rm USp}(2)_R \times {\rm U}(2)} \\
	\hline \hline \rm{stack} & N & (n^1,l^1)\times (n^2,l^2)\times (n^3,l^3) & n_{\yng(2)} & n_{\yng(1,1)_{}} & b & b' & c & d & d' & 1 & 2 & 3 & 4 \\
	\hline
	a   & 8   & (-1, -3)\times (0, -1)\times (-1, -1) & -2    & 2     & -3       & 0    & 3     & 4     & 4     & 0     & 0     & 0    & 0 \\
    b   & 4   & (0, 1)\times (3, -4)\times (1, -1)    & 1     & -1    & \text{-} & \text{-}  & -4  & 4     & -10   & 0     & 0     & 0    & 0 \\
    c   & 2   & (1, 0)\times (1, 0)\times (2, 0)      & 0     & 0     & \text{-} & \text{-}     & \text{-}    & 2     & -2    & 0     & 0     & 0    & 0 \\
    d   & 4   & (-1, 1)\times (1, -2)\times (-3, -1)  & -2    & -22   & \text{-} & \text{-}  & \text{-}  & \text{-}  & \text{-}  & 0     & 0     & 0    & 0 \\
	\hline
          &     &                     &  \multicolumn{11}{c|}{\beta^g_{d}=8, \quad  x_A=\frac{3}{4}x_B=\frac{1}{3}x_C=\frac{1}{24}x_D} \\
          &     &                     &  \multicolumn{11}{c|}{\chi_1=\sqrt{6}/18, \quad \chi_2=\sqrt{6}/8, \quad \chi_3= 2\sqrt{6}} \\
	\hline
\end{array}$
\end{tabular}\end{table}

\begin{table}[htb]  
\centering \footnotesize
\caption{D6-brane configurations and intersection numbers of Model~\hyperref[model2.1]{2}, and its gauge coupling relation is $g^2_a=\frac{11}{36}\, g^2_b=\frac{1}{3}\, g_c^2= \frac{3}{5}\,\left(\frac{5}{3}\,g^2_Y\right)= \frac{2 \sqrt{2}}{9} 11^{3/4}\, \pi \,e^{\phi_4}$.\\} \label{model2.1}
\begin{tabular}{c}$\begin{array}{|c|c|c |r|r|r| r|r|r| r|r|r| r|r|r|}
	\hline 	\multicolumn{3}{|c|}{\rm{Model~\hyperref[model2.1]{2}}}  & \multicolumn{11}{c|}{{\rm U}(4)\times {\rm USp}(2)_L \times {\rm U}(2)_R \times {\rm U}(2) \times {\rm USp}(2)}\\
	\hline \hline \rm{stack} & N & (n^1,l^1)\times (n^2,l^2)\times (n^3,l^3) & n_{\yng(2)} & n_{\yng(1,1)_{}} & b & c & c'& d & d' & 1 & 2 & 3 & 4 \\
	\hline
a     & 8   & (-1, -3)\times (-1, 0)\times (1, -1) & -2    & 2     & 3        & 0        & -3       & 4     & 4           & 0     & 0     & 0     & 0 \\
b     & 2   & (1, 0)\times (0, 1)\times (0, -2)    & 0     & 0     & \text{-} & -1       & 1        & 1     & -1          & 0     & 0     & 0     & 0 \\
c     & 4   & (0, 1)\times (1, -3)\times (1, -1)   & 2     & -2    & \text{-} & \text{-} & \text{-} & 4     & -4          & -3    & 0     & 0     & 0 \\
d     & 4   & (-1, 1)\times (-1, -1)\times (1, -3) & 0     & -12   & \text{-} & \text{-} & \text{-} & \text{-} & \text{-} & -3    & 0     & 0     & 0 \\
\hline
1       & 2   & (1, 0)\times (1, 0)\times (2, 0) &  \multicolumn{11}{c|}{\beta^g_{d}=7,~ \beta^g_1=0, \quad  x_A=\frac{1}{3}x_B=\frac{1}{33}x_C=\frac{1}{9}x_D} \\
        &     &                                  &  \multicolumn{11}{c|}{\chi_1=\sqrt{11}/33, \quad \chi_2=\sqrt{11}/3, \quad \chi_3= 2\sqrt{11}/11}  \\
	\hline
\end{array}$
\end{tabular}\end{table}

\begin{table}[htb]  
\centering \footnotesize
\caption{D6-brane configurations and intersection numbers of Model~\hyperref[model2.2]{3}, and its gauge coupling relation is $g^2_a=\frac{5}{18}\, g^2_b=\frac{1}{3}\, g_c^2= \frac{3}{5}\,\left(\frac{5}{3}\,g^2_Y\right)= \frac{4 \sqrt{2}}{9} 5^{3/4} \, \pi \,e^{\phi_4}$.\\} \label{model2.2}
\begin{tabular}{c}$\begin{array}{|c|c|c |r|r|r| r|r|r| r|r|r| r|r|r|}
	\hline 	\multicolumn{3}{|c|}{\rm{Model~\hyperref[model2.2]{3}}}  & \multicolumn{11}{c|}{{\rm U}(4)\times {\rm USp}(2)_L \times {\rm U}(2)_R \times {\rm U}(2) \times {\rm USp}(2)}\\
	\hline \hline \rm{stack} & N & (n^1,l^1)\times (n^2,l^2)\times (n^3,l^3) & n_{\yng(2)} & n_{\yng(1,1)_{}} & b & c & c'& d & d' & 1 & 2 & 3 & 4 \\
	\hline
a     & 8     & (-1, 0)\times (-1, -3)\times (1, -3) & 0     & 0       & 3     & -3    & 0     & 0     & 0     & 0     & 0     & 0     & 0 \\
b     & 2     & (0, 1)\times (1, 0)\times (0, -2) & 0     & 0         & \text{-}    & 1     & -1    & 6     & -6    & 0     & 0     & 0     & 0 \\
c     & 4     & (1, 1)\times (0, 1)\times (-1, -3) & -2    & 2         & \text{-}    & \text{-}    & \text{-}    & 3     & 0     & 0     & 0     & 0     & 0 \\
d     & 4     & (2, 1)\times (1, -3)\times (1, -3) & -16   & -56      & \text{-}    & \text{-}    & \text{-}    & \text{-}    & \text{-}    & 0     & 1     & 0     & 0 \\
\hline
2     & 2     & (1, 0)\times (0, -1)\times (0, 2) &  \multicolumn{11}{c|}{\beta^g_{d}=3,~ \beta^g_2=-5, \quad  x_A=\frac{1}{45}x_B=\frac{1}{3}x_C=\frac{1}{3}x_D} \\
      &       &                                   &  \multicolumn{11}{c|}{\chi_1=\sqrt{5}, \quad \chi_2=\sqrt{5}/15, \quad \chi_3= 2\sqrt{5}/15}  \\
	\hline
\end{array}$
\end{tabular}\end{table}

\begin{table}[htb]  
\centering \footnotesize
\caption{D6-brane configurations and intersection numbers of Model~\hyperref[model2.2dual]{3-dual}, and its gauge coupling relation is $g^2_a=\frac{1}{3}\, g^2_b=\frac{5}{18}\, g_c^2= \frac{17}{30}\,\left(\frac{5}{3}\,g^2_Y\right)= \frac{4 \sqrt{2}}{9} 5^{3/4} \, \pi \,e^{\phi_4}$.\\} \label{model2.2dual}
\begin{tabular}{c}$\begin{array}{|c|c|c |r|r|r| r|r|r| r|r|r| r|r|r|}
	\hline 	\multicolumn{3}{|c|}{\text{Model~\hyperref[model2.2dual]{3-dual}}}  & \multicolumn{11}{c|}{{\rm U}(4)\times {\rm U}(2)_L \times {\rm USp}(2)_R \times {\rm U}(2) \times {\rm USp}(2)}\\
	\hline \hline \rm{stack} & N & (n^1,l^1)\times (n^2,l^2)\times (n^3,l^3) & n_{\yng(2)} & n_{\yng(1,1)_{}} & b & b' & c & d & d' & 1 & 2 & 3 & 4 \\
	\hline
a     & 8     & (-1, 0)\times (-1, -3)\times (1, -3) & 0     & 0       & -3     & 0    & 3    & 0     & 0     & 0     & 0     & 0     & 0 \\
b     & 4     & (1, 1)\times (0, 1)\times (-1, -3)   & -2    & 2       & \text{-}    & \text{-}    & -1    & 3     & 0     & 0     & 0     & 0     & 0 \\
c     & 2     & (0, 1)\times (1, 0)\times (0, -2)    & 0     & 0       & \text{-}    & \text{-}     & \text{-}    & 6     & -6    & 0     & 0     & 0     & 0 \\
d     & 4     & (2, 1)\times (1, -3)\times (1, -3)   & -16   & -56     & \text{-}    & \text{-}    & \text{-}    & \text{-}    & \text{-}    & 0     & 1     & 0     & 0 \\
\hline
2     & 2     & (1, 0)\times (0, -1)\times (0, 2) &  \multicolumn{11}{c|}{\beta^g_{d}=3,~ \beta^g_2=-5, \quad  x_A=\frac{1}{45}x_B=\frac{1}{3}x_C=\frac{1}{3}x_D} \\
      &       &                                   &  \multicolumn{11}{c|}{\chi_1=\sqrt{5}, \quad \chi_2=\sqrt{5}/15, \quad \chi_3= 2\sqrt{5}/15}  \\
	\hline
\end{array}$
\end{tabular}\end{table}

\begin{table}[htb] 
\centering \footnotesize
\caption{D6-brane configurations and intersection numbers of Model~\hyperref[model4]{4}, and its gauge coupling relation is $g^2_a=\frac{11}{18}\, g^2_b=\frac{2}{3}\, g_c^2= \frac{4}{5}\,\left(\frac{5}{3}\,g^2_Y\right)= \frac{4}{9}\, 11^{3/4}\, \pi \,e^{\phi_4}$.\\} \label{model4}
\begin{tabular}{c}$\begin{array}{|c|c|c |r|r|r| r|r|r| r|r|r| r|r|r|}
\hline 	\multicolumn{3}{|c|}{\rm{Model~\hyperref[model4]{4}}}  & \multicolumn{11}{c|}{{\rm U}(4)\times {\rm USp}(2)_L \times {\rm U}(2)_R \times {\rm U}(2) \times {\rm USp}(4)}\\
\hline \hline \rm{stack} & N & (n^1,l^1)\times (n^2,l^2)\times (n^3,l^3) & n_{\yng(2)} & n_{\yng(1,1)_{}} & b  & c & c'& d & d' & 1 & 2 & 3 & 4 \\
\hline
a     & 8     & (-1, 0)\times (-1, -3)\times (1, -1) & -2    & 2      & 3     & -3    & 0     & 4     & 4     & 0     & 0     & 0     & 0 \\
b     & 2     & (0, 1)\times (1, 0)\times (0, -2) & 0     & 0         & \text{-}    & 2     & -2    & 2     & -2    & 0     & 0     & 0     & 0 \\
c     & 4     & (2, 3)\times (0, 1)\times (-1, -1) & -1    & 1         & \text{-}    & \text{-}    & \text{-}    & 8     & -8    & 0     & 0     & 0     & 0 \\
d     & 4     & (2, 1)\times (1, -1)\times (1, -3) & -2    & -22      & \text{-}    & \text{-}    & \text{-}    & \text{-}    & \text{-}    & 0     & 1     & 0     & 0 \\
\hline
2     & 4     & (1, 0)\times (0, -1)\times (0, 2) &  \multicolumn{11}{c|}{\beta^g_{d}=12,~ \beta^g_2=-5, \quad  x_A=\frac{1}{33}x_B=\frac{2}{3}x_C=\frac{2}{9}x_D} \\
      &       &                      &   \multicolumn{11}{c|}{\chi_1=2\sqrt{11}/3, \quad \chi_2=\sqrt{11}/33, \quad \chi_3= 2\sqrt{11}/11}  \\
\hline
\end{array}$
\end{tabular}\end{table}

\begin{table}[htb] 
\centering \footnotesize
\caption{D6-brane configurations and intersection numbers of Model~\hyperref[model6]{5}, and its gauge coupling relation is $g^2_a=\frac{7}{22}\, g^2_b=\frac{1}{3}\, g_c^2= \frac{3}{5}\,\left(\frac{5}{3}\,g^2_Y\right)= \frac{4 \sqrt{2}}{11}\frac{7^{3/4}}{\sqrt[4]{3}}\, \pi \,e^{\phi_4}$.\\} \label{model6}
\begin{tabular}{c}$\begin{array}{|c|c|c |r|r|r| r|r|r| r|r|r| r|r|r|}
\hline 	\multicolumn{3}{|c|}{\rm{Model~\hyperref[model6]{5}}}  & \multicolumn{11}{c|}{{\rm U}(4)\times {\rm USp}(2)_L \times {\rm U}(2)_R \times {\rm U}(2)\times {\rm USp}(6)}\\
\hline \hline \rm{stack} & N & (n^1,l^1)\times (n^2,l^2)\times (n^3,l^3) & n_{\yng(2)} & n_{\yng(1,1)_{}} & b & c & c'& d & d' & 1 & 2 & 3 & 4 \\
\hline
a     & 8     & (1, 3)\times (1, 0)\times (1, -1) & -2    & 2       & 3     & 0     & -3    & 4     & 4     & 0     & 0     & 0     & 0 \\
b     & 2     & (1, 0)\times (0, 1)\times (0, -2) & 0     & 0       & \text{-}    & -1    & 1     & 2     & -2    & 0     & 0     & 0     & 0 \\
c     & 4     & (0, -1)\times (-1, 3)\times (1, -1) & 2     & -2       & \text{-}    & \text{-}    & \text{-}    & 7     & -10   & 0     & 0     & 0     & 0 \\
d     & 4     & (1, -1)\times (2, 1)\times (1, -3) & -2    & -22      & \text{-}    & \text{-}    & \text{-}    & \text{-}    & \text{-}    & 0     & 0     & 1     & 0 \\
\hline
3     & 6     & (0, -1)\times (1, 0)\times (0, 2) &  \multicolumn{11}{c|}{\beta^g_{d}=11,~ \beta^g_3=-5, \quad  x_A=\frac{1}{3}x_B=\frac{1}{63}x_C=\frac{1}{9}x_D} \\
      &       &                                   &   \multicolumn{11}{c|}{\chi_1=\sqrt{21}/63, \quad \chi_2=\sqrt{21}/3, \quad \chi_3= 2\sqrt{2}/21}  \\
\hline
\end{array}$
\end{tabular}\end{table}

\FloatBarrier
 
\bibliographystyle{apsrev4-2}

\begin{thebibliography}{51}%
\makeatletter
\providecommand \@ifxundefined [1]{%
 \@ifx{#1\undefined}
}%
\providecommand \@ifnum [1]{%
 \ifnum #1\expandafter \@firstoftwo
 \else \expandafter \@secondoftwo
 \fi
}%
\providecommand \@ifx [1]{%
 \ifx #1\expandafter \@firstoftwo
 \else \expandafter \@secondoftwo
 \fi
}%
\providecommand \natexlab [1]{#1}%
\providecommand \enquote  [1]{``#1''}%
\providecommand \bibnamefont  [1]{#1}%
\providecommand \bibfnamefont [1]{#1}%
\providecommand \citenamefont [1]{#1}%
\providecommand \href@noop [0]{\@secondoftwo}%
\providecommand \href [0]{\begingroup \@sanitize@url \@href}%
\providecommand \@href[1]{\@@startlink{#1}\@@href}%
\providecommand \@@href[1]{\endgroup#1\@@endlink}%
\providecommand \@sanitize@url [0]{\catcode `\\12\catcode `\$12\catcode
  `\&12\catcode `\#12\catcode `\^12\catcode `\_12\catcode `\%12\relax}%
\providecommand \@@startlink[1]{}%
\providecommand \@@endlink[0]{}%
\providecommand \url  [0]{\begingroup\@sanitize@url \@url }%
\providecommand \@url [1]{\endgroup\@href {#1}{\urlprefix }}%
\providecommand \urlprefix  [0]{URL }%
\providecommand \Eprint [0]{\href }%
\providecommand \doibase [0]{https://doi.org/}%
\providecommand \selectlanguage [0]{\@gobble}%
\providecommand \bibinfo  [0]{\@secondoftwo}%
\providecommand \bibfield  [0]{\@secondoftwo}%
\providecommand \translation [1]{[#1]}%
\providecommand \BibitemOpen [0]{}%
\providecommand \bibitemStop [0]{}%
\providecommand \bibitemNoStop [0]{.\EOS\space}%
\providecommand \EOS [0]{\spacefactor3000\relax}%
\providecommand \BibitemShut  [1]{\csname bibitem#1\endcsname}%
\let\auto@bib@innerbib\@empty
\bibitem [{\citenamefont {Chamoun}\ \emph {et~al.}(2004)\citenamefont
  {Chamoun}, \citenamefont {Khalil},\ and\ \citenamefont
  {Lashin}}]{Chamoun:2003pf}%
  \BibitemOpen
  \bibfield  {author} {\bibinfo {author} {\bibfnamefont {N.}~\bibnamefont
  {Chamoun}}, \bibinfo {author} {\bibfnamefont {S.}~\bibnamefont {Khalil}},\
  and\ \bibinfo {author} {\bibfnamefont {E.}~\bibnamefont {Lashin}},\ }\href
  {https://doi.org/10.1103/PhysRevD.69.095011} {\bibfield  {journal} {\bibinfo
  {journal} {Phys. Rev. D}\ }\textbf {\bibinfo {volume} {69}},\ \bibinfo
  {pages} {095011} (\bibinfo {year} {2004})},\ \Eprint
  {https://arxiv.org/abs/hep-ph/0309169} {arXiv:hep-ph/0309169} \BibitemShut
  {NoStop}%
\bibitem [{\citenamefont {Higaki}\ \emph {et~al.}(2005)\citenamefont {Higaki},
  \citenamefont {Kitazawa}, \citenamefont {Kobayashi},\ and\ \citenamefont
  {Takahashi}}]{Higaki:2005ie}%
  \BibitemOpen
  \bibfield  {author} {\bibinfo {author} {\bibfnamefont {T.}~\bibnamefont
  {Higaki}}, \bibinfo {author} {\bibfnamefont {N.}~\bibnamefont {Kitazawa}},
  \bibinfo {author} {\bibfnamefont {T.}~\bibnamefont {Kobayashi}},\ and\
  \bibinfo {author} {\bibfnamefont {K.-j.}\ \bibnamefont {Takahashi}},\ }\href
  {https://doi.org/10.1103/PhysRevD.72.086003} {\bibfield  {journal} {\bibinfo
  {journal} {Phys. Rev. D}\ }\textbf {\bibinfo {volume} {72}},\ \bibinfo
  {pages} {086003} (\bibinfo {year} {2005})},\ \Eprint
  {https://arxiv.org/abs/hep-th/0504019} {arXiv:hep-th/0504019} \BibitemShut
  {NoStop}%
\bibitem [{\citenamefont {Aldazabal}\ \emph {et~al.}(2000)\citenamefont
  {Aldazabal}, \citenamefont {Ibanez}, \citenamefont {Quevedo},\ and\
  \citenamefont {Uranga}}]{Aldazabal:2000sa}%
  \BibitemOpen
  \bibfield  {author} {\bibinfo {author} {\bibfnamefont {G.}~\bibnamefont
  {Aldazabal}}, \bibinfo {author} {\bibfnamefont {L.~E.}\ \bibnamefont
  {Ibanez}}, \bibinfo {author} {\bibfnamefont {F.}~\bibnamefont {Quevedo}},\
  and\ \bibinfo {author} {\bibfnamefont {A.~M.}\ \bibnamefont {Uranga}},\
  }\href {https://doi.org/10.1088/1126-6708/2000/08/002} {\bibfield  {journal}
  {\bibinfo  {journal} {JHEP}\ }\textbf {\bibinfo {volume} {08}},\ \bibinfo
  {pages} {002}},\ \Eprint {https://arxiv.org/abs/hep-th/0005067}
  {arXiv:hep-th/0005067} \BibitemShut {NoStop}%
\bibitem [{\citenamefont {Shiu}\ and\ \citenamefont {Tye}(1998)}]{Shiu:1998pa}%
  \BibitemOpen
  \bibfield  {author} {\bibinfo {author} {\bibfnamefont {G.}~\bibnamefont
  {Shiu}}\ and\ \bibinfo {author} {\bibfnamefont {S.~H.~H.}\ \bibnamefont
  {Tye}},\ }\href {https://doi.org/10.1103/PhysRevD.58.106007} {\bibfield
  {journal} {\bibinfo  {journal} {Phys. Rev. D}\ }\textbf {\bibinfo {volume}
  {58}},\ \bibinfo {pages} {106007} (\bibinfo {year} {1998})},\ \Eprint
  {https://arxiv.org/abs/hep-th/9805157} {arXiv:hep-th/9805157} \BibitemShut
  {NoStop}%
\bibitem [{\citenamefont {Cvetic}\ \emph
  {et~al.}(2001{\natexlab{a}})\citenamefont {Cvetic}, \citenamefont {Shiu},\
  and\ \citenamefont {Uranga}}]{Cvetic:2001tj}%
  \BibitemOpen
  \bibfield  {author} {\bibinfo {author} {\bibfnamefont {M.}~\bibnamefont
  {Cvetic}}, \bibinfo {author} {\bibfnamefont {G.}~\bibnamefont {Shiu}},\ and\
  \bibinfo {author} {\bibfnamefont {A.~M.}\ \bibnamefont {Uranga}},\ }\href
  {https://doi.org/10.1103/PhysRevLett.87.201801} {\bibfield  {journal}
  {\bibinfo  {journal} {Phys. Rev. Lett.}\ }\textbf {\bibinfo {volume} {87}},\
  \bibinfo {pages} {201801} (\bibinfo {year} {2001}{\natexlab{a}})},\ \Eprint
  {https://arxiv.org/abs/hep-th/0107143} {arXiv:hep-th/0107143} \BibitemShut
  {NoStop}%
\bibitem [{\citenamefont {Cvetic}\ \emph {et~al.}(2004)\citenamefont {Cvetic},
  \citenamefont {Li},\ and\ \citenamefont {Liu}}]{Cvetic:2004ui}%
  \BibitemOpen
  \bibfield  {author} {\bibinfo {author} {\bibfnamefont {M.}~\bibnamefont
  {Cvetic}}, \bibinfo {author} {\bibfnamefont {T.}~\bibnamefont {Li}},\ and\
  \bibinfo {author} {\bibfnamefont {T.}~\bibnamefont {Liu}},\ }\href
  {https://doi.org/10.1016/j.nuclphysb.2004.07.036} {\bibfield  {journal}
  {\bibinfo  {journal} {Nucl. Phys. B}\ }\textbf {\bibinfo {volume} {698}},\
  \bibinfo {pages} {163} (\bibinfo {year} {2004})},\ \Eprint
  {https://arxiv.org/abs/hep-th/0403061} {arXiv:hep-th/0403061} \BibitemShut
  {NoStop}%
\bibitem [{\citenamefont {Li}\ \emph {et~al.}(2021{\natexlab{a}})\citenamefont
  {Li}, \citenamefont {Mansha},\ and\ \citenamefont {Sun}}]{Li:2019nvi}%
  \BibitemOpen
  \bibfield  {author} {\bibinfo {author} {\bibfnamefont {T.}~\bibnamefont
  {Li}}, \bibinfo {author} {\bibfnamefont {A.}~\bibnamefont {Mansha}},\ and\
  \bibinfo {author} {\bibfnamefont {R.}~\bibnamefont {Sun}},\ }\href
  {https://doi.org/10.1140/epjc/s10052-021-08839-w} {\bibfield  {journal}
  {\bibinfo  {journal} {Eur. Phys. J. C}\ }\textbf {\bibinfo {volume} {81}},\
  \bibinfo {pages} {82} (\bibinfo {year} {2021}{\natexlab{a}})},\ \Eprint
  {https://arxiv.org/abs/1910.04530} {arXiv:1910.04530 [hep-th]} \BibitemShut
  {NoStop}%
\bibitem [{\citenamefont {Chen}\ \emph
  {et~al.}(2008{\natexlab{a}})\citenamefont {Chen}, \citenamefont {Li},
  \citenamefont {Mayes},\ and\ \citenamefont {Nanopoulos}}]{Chen:2007px}%
  \BibitemOpen
  \bibfield  {author} {\bibinfo {author} {\bibfnamefont {C.-M.}\ \bibnamefont
  {Chen}}, \bibinfo {author} {\bibfnamefont {T.}~\bibnamefont {Li}}, \bibinfo
  {author} {\bibfnamefont {V.~E.}\ \bibnamefont {Mayes}},\ and\ \bibinfo
  {author} {\bibfnamefont {D.~V.}\ \bibnamefont {Nanopoulos}},\ }\href
  {https://doi.org/10.1016/j.physletb.2008.06.024} {\bibfield  {journal}
  {\bibinfo  {journal} {Phys. Lett. B}\ }\textbf {\bibinfo {volume} {665}},\
  \bibinfo {pages} {267} (\bibinfo {year} {2008}{\natexlab{a}})},\ \Eprint
  {https://arxiv.org/abs/hep-th/0703280} {arXiv:hep-th/0703280} \BibitemShut
  {NoStop}%
\bibitem [{\citenamefont {Chen}\ \emph
  {et~al.}(2008{\natexlab{b}})\citenamefont {Chen}, \citenamefont {Li},
  \citenamefont {Mayes},\ and\ \citenamefont {Nanopoulos}}]{Chen:2007zu}%
  \BibitemOpen
  \bibfield  {author} {\bibinfo {author} {\bibfnamefont {C.-M.}\ \bibnamefont
  {Chen}}, \bibinfo {author} {\bibfnamefont {T.}~\bibnamefont {Li}}, \bibinfo
  {author} {\bibfnamefont {V.~E.}\ \bibnamefont {Mayes}},\ and\ \bibinfo
  {author} {\bibfnamefont {D.~V.}\ \bibnamefont {Nanopoulos}},\ }\href
  {https://doi.org/10.1103/PhysRevD.77.125023} {\bibfield  {journal} {\bibinfo
  {journal} {Phys. Rev. D}\ }\textbf {\bibinfo {volume} {77}},\ \bibinfo
  {pages} {125023} (\bibinfo {year} {2008}{\natexlab{b}})},\ \Eprint
  {https://arxiv.org/abs/0711.0396} {arXiv:0711.0396 [hep-ph]} \BibitemShut
  {NoStop}%
\bibitem [{\citenamefont {Sabir}\ \emph {et~al.}(2022)\citenamefont {Sabir},
  \citenamefont {Li}, \citenamefont {Mansha},\ and\ \citenamefont
  {Wang}}]{Sabir:2022hko}%
  \BibitemOpen
  \bibfield  {author} {\bibinfo {author} {\bibfnamefont {M.}~\bibnamefont
  {Sabir}}, \bibinfo {author} {\bibfnamefont {T.}~\bibnamefont {Li}}, \bibinfo
  {author} {\bibfnamefont {A.}~\bibnamefont {Mansha}},\ and\ \bibinfo {author}
  {\bibfnamefont {X.-C.}\ \bibnamefont {Wang}},\ }\href
  {https://doi.org/10.1007/JHEP04(2022)089} {\bibfield  {journal} {\bibinfo
  {journal} {JHEP}\ }\textbf {\bibinfo {volume} {04}},\ \bibinfo {pages}
  {089}},\ \Eprint {https://arxiv.org/abs/2202.07048} {arXiv:2202.07048
  [hep-th]} \BibitemShut {NoStop}%
\bibitem [{\citenamefont {He}\ \emph {et~al.}(2022{\natexlab{a}})\citenamefont
  {He}, \citenamefont {Li}, \citenamefont {Sun},\ and\ \citenamefont
  {Wu}}]{He:2021kbj}%
  \BibitemOpen
  \bibfield  {author} {\bibinfo {author} {\bibfnamefont {W.}~\bibnamefont
  {He}}, \bibinfo {author} {\bibfnamefont {T.}~\bibnamefont {Li}}, \bibinfo
  {author} {\bibfnamefont {R.}~\bibnamefont {Sun}},\ and\ \bibinfo {author}
  {\bibfnamefont {L.}~\bibnamefont {Wu}},\ }\href
  {https://doi.org/10.1140/epjc/s10052-022-10663-9} {\bibfield  {journal}
  {\bibinfo  {journal} {Eur. Phys. J. C}\ }\textbf {\bibinfo {volume} {82}},\
  \bibinfo {pages} {710} (\bibinfo {year} {2022}{\natexlab{a}})},\ \Eprint
  {https://arxiv.org/abs/2112.09630} {arXiv:2112.09630 [hep-th]} \BibitemShut
  {NoStop}%
\bibitem [{\citenamefont {He}\ \emph {et~al.}(2022{\natexlab{b}})\citenamefont
  {He}, \citenamefont {Li},\ and\ \citenamefont {Sun}}]{He:2021gug}%
  \BibitemOpen
  \bibfield  {author} {\bibinfo {author} {\bibfnamefont {W.}~\bibnamefont
  {He}}, \bibinfo {author} {\bibfnamefont {T.}~\bibnamefont {Li}},\ and\
  \bibinfo {author} {\bibfnamefont {R.}~\bibnamefont {Sun}},\ }\href
  {https://doi.org/10.1007/JHEP08(2022)044} {\bibfield  {journal} {\bibinfo
  {journal} {JHEP}\ }\textbf {\bibinfo {volume} {08}},\ \bibinfo {pages}
  {044}},\ \Eprint {https://arxiv.org/abs/2112.09632} {arXiv:2112.09632
  [hep-th]} \BibitemShut {NoStop}%
\bibitem [{\citenamefont {Li}\ \emph {et~al.}(2021{\natexlab{b}})\citenamefont
  {Li}, \citenamefont {Mansha}, \citenamefont {Sun}, \citenamefont {Wu},\ and\
  \citenamefont {He}}]{Li:2021pxo}%
  \BibitemOpen
  \bibfield  {author} {\bibinfo {author} {\bibfnamefont {T.}~\bibnamefont
  {Li}}, \bibinfo {author} {\bibfnamefont {A.}~\bibnamefont {Mansha}}, \bibinfo
  {author} {\bibfnamefont {R.}~\bibnamefont {Sun}}, \bibinfo {author}
  {\bibfnamefont {L.}~\bibnamefont {Wu}},\ and\ \bibinfo {author}
  {\bibfnamefont {W.}~\bibnamefont {He}},\ }\href
  {https://doi.org/10.1103/PhysRevD.104.046018} {\bibfield  {journal} {\bibinfo
   {journal} {Phys. Rev. D}\ }\textbf {\bibinfo {volume} {104}},\ \bibinfo
  {pages} {046018} (\bibinfo {year} {2021}{\natexlab{b}})}\BibitemShut
  {NoStop}%
\bibitem [{\citenamefont {Cvetic}\ \emph {et~al.}(2005)\citenamefont {Cvetic},
  \citenamefont {Langacker}, \citenamefont {Li},\ and\ \citenamefont
  {Liu}}]{Cvetic:2004nk}%
  \BibitemOpen
  \bibfield  {author} {\bibinfo {author} {\bibfnamefont {M.}~\bibnamefont
  {Cvetic}}, \bibinfo {author} {\bibfnamefont {P.}~\bibnamefont {Langacker}},
  \bibinfo {author} {\bibfnamefont {T.-j.}\ \bibnamefont {Li}},\ and\ \bibinfo
  {author} {\bibfnamefont {T.}~\bibnamefont {Liu}},\ }\href
  {https://doi.org/10.1016/j.nuclphysb.2004.12.028} {\bibfield  {journal}
  {\bibinfo  {journal} {Nucl. Phys. B}\ }\textbf {\bibinfo {volume} {709}},\
  \bibinfo {pages} {241} (\bibinfo {year} {2005})},\ \Eprint
  {https://arxiv.org/abs/hep-th/0407178} {arXiv:hep-th/0407178} \BibitemShut
  {NoStop}%
\bibitem [{\citenamefont {Gimon}\ and\ \citenamefont
  {Polchinski}(1996)}]{Gimon:1996rq}%
  \BibitemOpen
  \bibfield  {author} {\bibinfo {author} {\bibfnamefont {E.~G.}\ \bibnamefont
  {Gimon}}\ and\ \bibinfo {author} {\bibfnamefont {J.}~\bibnamefont
  {Polchinski}},\ }\href {https://doi.org/10.1103/PhysRevD.54.1667} {\bibfield
  {journal} {\bibinfo  {journal} {Phys. Rev. D}\ }\textbf {\bibinfo {volume}
  {54}},\ \bibinfo {pages} {1667} (\bibinfo {year} {1996})},\ \Eprint
  {https://arxiv.org/abs/hep-th/9601038} {arXiv:hep-th/9601038} \BibitemShut
  {NoStop}%
\bibitem [{\citenamefont {Green}\ and\ \citenamefont
  {Schwarz}(1984)}]{Green:1984sg}%
  \BibitemOpen
  \bibfield  {author} {\bibinfo {author} {\bibfnamefont {M.~B.}\ \bibnamefont
  {Green}}\ and\ \bibinfo {author} {\bibfnamefont {J.~H.}\ \bibnamefont
  {Schwarz}},\ }\href {https://doi.org/10.1016/0370-2693(84)91565-X} {\bibfield
   {journal} {\bibinfo  {journal} {Phys. Lett. B}\ }\textbf {\bibinfo {volume}
  {149}},\ \bibinfo {pages} {117} (\bibinfo {year} {1984})}\BibitemShut
  {NoStop}%
\bibitem [{\citenamefont {Witten}(1998)}]{Witten:1998cd}%
  \BibitemOpen
  \bibfield  {author} {\bibinfo {author} {\bibfnamefont {E.}~\bibnamefont
  {Witten}},\ }\href {https://doi.org/10.1088/1126-6708/1998/12/019} {\bibfield
   {journal} {\bibinfo  {journal} {JHEP}\ }\textbf {\bibinfo {volume} {12}},\
  \bibinfo {pages} {019}},\ \Eprint {https://arxiv.org/abs/hep-th/9810188}
  {arXiv:hep-th/9810188} \BibitemShut {NoStop}%
\bibitem [{\citenamefont {Cascales}\ and\ \citenamefont
  {Uranga}(2003)}]{Cascales:2003zp}%
  \BibitemOpen
  \bibfield  {author} {\bibinfo {author} {\bibfnamefont {J.~F.~G.}\
  \bibnamefont {Cascales}}\ and\ \bibinfo {author} {\bibfnamefont {A.~M.}\
  \bibnamefont {Uranga}},\ }\href
  {https://doi.org/10.1088/1126-6708/2003/05/011} {\bibfield  {journal}
  {\bibinfo  {journal} {JHEP}\ }\textbf {\bibinfo {volume} {05}},\ \bibinfo
  {pages} {011}},\ \Eprint {https://arxiv.org/abs/hep-th/0303024}
  {arXiv:hep-th/0303024} \BibitemShut {NoStop}%
\bibitem [{\citenamefont {Marchesano}\ and\ \citenamefont
  {Shiu}(2005)}]{Marchesano:2004yq}%
  \BibitemOpen
  \bibfield  {author} {\bibinfo {author} {\bibfnamefont {F.}~\bibnamefont
  {Marchesano}}\ and\ \bibinfo {author} {\bibfnamefont {G.}~\bibnamefont
  {Shiu}},\ }\href {https://doi.org/10.1103/PhysRevD.71.011701} {\bibfield
  {journal} {\bibinfo  {journal} {Phys. Rev. D}\ }\textbf {\bibinfo {volume}
  {71}},\ \bibinfo {pages} {011701} (\bibinfo {year} {2005})},\ \Eprint
  {https://arxiv.org/abs/hep-th/0408059} {arXiv:hep-th/0408059} \BibitemShut
  {NoStop}%
\bibitem [{\citenamefont {Marchesano}\ and\ \citenamefont
  {Shiu}(2004)}]{Marchesano:2004xz}%
  \BibitemOpen
  \bibfield  {author} {\bibinfo {author} {\bibfnamefont {F.}~\bibnamefont
  {Marchesano}}\ and\ \bibinfo {author} {\bibfnamefont {G.}~\bibnamefont
  {Shiu}},\ }\href {https://doi.org/10.1088/1126-6708/2004/11/041} {\bibfield
  {journal} {\bibinfo  {journal} {JHEP}\ }\textbf {\bibinfo {volume} {11}},\
  \bibinfo {pages} {041}},\ \Eprint {https://arxiv.org/abs/hep-th/0409132}
  {arXiv:hep-th/0409132} \BibitemShut {NoStop}%
\bibitem [{\citenamefont {Uranga}(2001)}]{Uranga:2000xp}%
  \BibitemOpen
  \bibfield  {author} {\bibinfo {author} {\bibfnamefont {A.~M.}\ \bibnamefont
  {Uranga}},\ }\href {https://doi.org/10.1016/S0550-3213(00)00787-2} {\bibfield
   {journal} {\bibinfo  {journal} {Nucl. Phys. B}\ }\textbf {\bibinfo {volume}
  {598}},\ \bibinfo {pages} {225} (\bibinfo {year} {2001})},\ \Eprint
  {https://arxiv.org/abs/hep-th/0011048} {arXiv:hep-th/0011048} \BibitemShut
  {NoStop}%
\bibitem [{\citenamefont {Ibanez}\ \emph {et~al.}(2001)\citenamefont {Ibanez},
  \citenamefont {Marchesano},\ and\ \citenamefont {Rabadan}}]{Ibanez:2001nd}%
  \BibitemOpen
  \bibfield  {author} {\bibinfo {author} {\bibfnamefont {L.~E.}\ \bibnamefont
  {Ibanez}}, \bibinfo {author} {\bibfnamefont {F.}~\bibnamefont {Marchesano}},\
  and\ \bibinfo {author} {\bibfnamefont {R.}~\bibnamefont {Rabadan}},\ }\href
  {https://doi.org/10.1088/1126-6708/2001/11/002} {\bibfield  {journal}
  {\bibinfo  {journal} {JHEP}\ }\textbf {\bibinfo {volume} {11}},\ \bibinfo
  {pages} {002}},\ \Eprint {https://arxiv.org/abs/hep-th/0105155}
  {arXiv:hep-th/0105155} \BibitemShut {NoStop}%
\bibitem [{\citenamefont {Cvetic}\ \emph
  {et~al.}(2001{\natexlab{b}})\citenamefont {Cvetic}, \citenamefont {Shiu},\
  and\ \citenamefont {Uranga}}]{Cvetic:2001nr}%
  \BibitemOpen
  \bibfield  {author} {\bibinfo {author} {\bibfnamefont {M.}~\bibnamefont
  {Cvetic}}, \bibinfo {author} {\bibfnamefont {G.}~\bibnamefont {Shiu}},\ and\
  \bibinfo {author} {\bibfnamefont {A.~M.}\ \bibnamefont {Uranga}},\ }\href
  {https://doi.org/10.1016/S0550-3213(01)00427-8} {\bibfield  {journal}
  {\bibinfo  {journal} {Nucl. Phys. B}\ }\textbf {\bibinfo {volume} {615}},\
  \bibinfo {pages} {3} (\bibinfo {year} {2001}{\natexlab{b}})},\ \Eprint
  {https://arxiv.org/abs/hep-th/0107166} {arXiv:hep-th/0107166} \BibitemShut
  {NoStop}%
\bibitem [{\citenamefont {Cvetic}\ \emph
  {et~al.}(2003{\natexlab{a}})\citenamefont {Cvetic}, \citenamefont
  {Papadimitriou},\ and\ \citenamefont {Shiu}}]{Cvetic:2002pj}%
  \BibitemOpen
  \bibfield  {author} {\bibinfo {author} {\bibfnamefont {M.}~\bibnamefont
  {Cvetic}}, \bibinfo {author} {\bibfnamefont {I.}~\bibnamefont
  {Papadimitriou}},\ and\ \bibinfo {author} {\bibfnamefont {G.}~\bibnamefont
  {Shiu}},\ }\href {https://doi.org/10.1016/j.nuclphysb.2004.06.041} {\bibfield
   {journal} {\bibinfo  {journal} {Nucl. Phys. B}\ }\textbf {\bibinfo {volume}
  {659}},\ \bibinfo {pages} {193} (\bibinfo {year} {2003}{\natexlab{a}})},\
  \bibinfo {note} {[Erratum: Nucl.Phys.B 696, 298--298 (2004)]},\ \Eprint
  {https://arxiv.org/abs/hep-th/0212177} {arXiv:hep-th/0212177} \BibitemShut
  {NoStop}%
\bibitem [{\citenamefont {Cvetic}\ and\ \citenamefont
  {Papadimitriou}(2003)}]{Cvetic:2003xs}%
  \BibitemOpen
  \bibfield  {author} {\bibinfo {author} {\bibfnamefont {M.}~\bibnamefont
  {Cvetic}}\ and\ \bibinfo {author} {\bibfnamefont {I.}~\bibnamefont
  {Papadimitriou}},\ }\href {https://doi.org/10.1103/PhysRevD.67.126006}
  {\bibfield  {journal} {\bibinfo  {journal} {Phys. Rev. D}\ }\textbf {\bibinfo
  {volume} {67}},\ \bibinfo {pages} {126006} (\bibinfo {year} {2003})},\
  \Eprint {https://arxiv.org/abs/hep-th/0303197} {arXiv:hep-th/0303197}
  \BibitemShut {NoStop}%
\bibitem [{\citenamefont {Aldazabal}\ \emph {et~al.}(2001)\citenamefont
  {Aldazabal}, \citenamefont {Franco}, \citenamefont {Ibanez}, \citenamefont
  {Rabadan},\ and\ \citenamefont {Uranga}}]{Aldazabal:2000dg}%
  \BibitemOpen
  \bibfield  {author} {\bibinfo {author} {\bibfnamefont {G.}~\bibnamefont
  {Aldazabal}}, \bibinfo {author} {\bibfnamefont {S.}~\bibnamefont {Franco}},
  \bibinfo {author} {\bibfnamefont {L.~E.}\ \bibnamefont {Ibanez}}, \bibinfo
  {author} {\bibfnamefont {R.}~\bibnamefont {Rabadan}},\ and\ \bibinfo {author}
  {\bibfnamefont {A.~M.}\ \bibnamefont {Uranga}},\ }\href
  {https://doi.org/10.1063/1.1376157} {\bibfield  {journal} {\bibinfo
  {journal} {J. Math. Phys.}\ }\textbf {\bibinfo {volume} {42}},\ \bibinfo
  {pages} {3103} (\bibinfo {year} {2001})},\ \Eprint
  {https://arxiv.org/abs/hep-th/0011073} {arXiv:hep-th/0011073} \BibitemShut
  {NoStop}%
\bibitem [{\citenamefont {Cvetic}\ \emph
  {et~al.}(2003{\natexlab{b}})\citenamefont {Cvetic}, \citenamefont
  {Langacker},\ and\ \citenamefont {Wang}}]{Cvetic:2003yd}%
  \BibitemOpen
  \bibfield  {author} {\bibinfo {author} {\bibfnamefont {M.}~\bibnamefont
  {Cvetic}}, \bibinfo {author} {\bibfnamefont {P.}~\bibnamefont {Langacker}},\
  and\ \bibinfo {author} {\bibfnamefont {J.}~\bibnamefont {Wang}},\ }\href
  {https://doi.org/10.1103/PhysRevD.68.046002} {\bibfield  {journal} {\bibinfo
  {journal} {Phys. Rev. D}\ }\textbf {\bibinfo {volume} {68}},\ \bibinfo
  {pages} {046002} (\bibinfo {year} {2003}{\natexlab{b}})},\ \Eprint
  {https://arxiv.org/abs/hep-th/0303208} {arXiv:hep-th/0303208} \BibitemShut
  {NoStop}%
\bibitem [{\citenamefont {Lust}\ \emph {et~al.}(2004)\citenamefont {Lust},
  \citenamefont {Mayr}, \citenamefont {Richter},\ and\ \citenamefont
  {Stieberger}}]{Lust:2004cx}%
  \BibitemOpen
  \bibfield  {author} {\bibinfo {author} {\bibfnamefont {D.}~\bibnamefont
  {Lust}}, \bibinfo {author} {\bibfnamefont {P.}~\bibnamefont {Mayr}}, \bibinfo
  {author} {\bibfnamefont {R.}~\bibnamefont {Richter}},\ and\ \bibinfo {author}
  {\bibfnamefont {S.}~\bibnamefont {Stieberger}},\ }\href
  {https://doi.org/10.1016/j.nuclphysb.2004.06.052} {\bibfield  {journal}
  {\bibinfo  {journal} {Nucl. Phys. B}\ }\textbf {\bibinfo {volume} {696}},\
  \bibinfo {pages} {205} (\bibinfo {year} {2004})},\ \Eprint
  {https://arxiv.org/abs/hep-th/0404134} {arXiv:hep-th/0404134} \BibitemShut
  {NoStop}%
\bibitem [{\citenamefont {Blumenhagen}\ \emph
  {et~al.}(2007{\natexlab{a}})\citenamefont {Blumenhagen}, \citenamefont
  {Kors}, \citenamefont {Lust},\ and\ \citenamefont
  {Stieberger}}]{Blumenhagen:2006ci}%
  \BibitemOpen
  \bibfield  {author} {\bibinfo {author} {\bibfnamefont {R.}~\bibnamefont
  {Blumenhagen}}, \bibinfo {author} {\bibfnamefont {B.}~\bibnamefont {Kors}},
  \bibinfo {author} {\bibfnamefont {D.}~\bibnamefont {Lust}},\ and\ \bibinfo
  {author} {\bibfnamefont {S.}~\bibnamefont {Stieberger}},\ }\href
  {https://doi.org/10.1016/j.physrep.2007.04.003} {\bibfield  {journal}
  {\bibinfo  {journal} {Phys. Rept.}\ }\textbf {\bibinfo {volume} {445}},\
  \bibinfo {pages} {1} (\bibinfo {year} {2007}{\natexlab{a}})},\ \Eprint
  {https://arxiv.org/abs/hep-th/0610327} {arXiv:hep-th/0610327} \BibitemShut
  {NoStop}%
\bibitem [{\citenamefont {Ginsparg}(1987)}]{Ginsparg:1987ee}%
  \BibitemOpen
  \bibfield  {author} {\bibinfo {author} {\bibfnamefont {P.~H.}\ \bibnamefont
  {Ginsparg}},\ }\href {https://doi.org/10.1016/0370-2693(87)90357-1}
  {\bibfield  {journal} {\bibinfo  {journal} {Phys. Lett. B}\ }\textbf
  {\bibinfo {volume} {197}},\ \bibinfo {pages} {139} (\bibinfo {year}
  {1987})}\BibitemShut {NoStop}%
\bibitem [{\citenamefont {Hamada}\ \emph {et~al.}(2015)\citenamefont {Hamada},
  \citenamefont {Kobayashi},\ and\ \citenamefont {Uemura}}]{Hamada:2014eia}%
  \BibitemOpen
  \bibfield  {author} {\bibinfo {author} {\bibfnamefont {Y.}~\bibnamefont
  {Hamada}}, \bibinfo {author} {\bibfnamefont {T.}~\bibnamefont {Kobayashi}},\
  and\ \bibinfo {author} {\bibfnamefont {S.}~\bibnamefont {Uemura}},\ }\href
  {https://doi.org/10.1016/j.nuclphysb.2015.06.005} {\bibfield  {journal}
  {\bibinfo  {journal} {Nucl. Phys. B}\ }\textbf {\bibinfo {volume} {897}},\
  \bibinfo {pages} {563} (\bibinfo {year} {2015})},\ \Eprint
  {https://arxiv.org/abs/1409.2740} {arXiv:1409.2740 [hep-th]} \BibitemShut
  {NoStop}%
\bibitem [{\citenamefont {Blumenhagen}\ \emph {et~al.}(2003)\citenamefont
  {Blumenhagen}, \citenamefont {Lust},\ and\ \citenamefont
  {Stieberger}}]{Blumenhagen:2003jy}%
  \BibitemOpen
  \bibfield  {author} {\bibinfo {author} {\bibfnamefont {R.}~\bibnamefont
  {Blumenhagen}}, \bibinfo {author} {\bibfnamefont {D.}~\bibnamefont {Lust}},\
  and\ \bibinfo {author} {\bibfnamefont {S.}~\bibnamefont {Stieberger}},\
  }\href {https://doi.org/10.1088/1126-6708/2003/07/036} {\bibfield  {journal}
  {\bibinfo  {journal} {JHEP}\ }\textbf {\bibinfo {volume} {07}},\ \bibinfo
  {pages} {036}},\ \Eprint {https://arxiv.org/abs/hep-th/0305146}
  {arXiv:hep-th/0305146} \BibitemShut {NoStop}%
\bibitem [{\citenamefont {Cvetic}\ \emph {et~al.}(2002)\citenamefont {Cvetic},
  \citenamefont {Langacker},\ and\ \citenamefont {Shiu}}]{Cvetic:2002qa}%
  \BibitemOpen
  \bibfield  {author} {\bibinfo {author} {\bibfnamefont {M.}~\bibnamefont
  {Cvetic}}, \bibinfo {author} {\bibfnamefont {P.}~\bibnamefont {Langacker}},\
  and\ \bibinfo {author} {\bibfnamefont {G.}~\bibnamefont {Shiu}},\ }\href
  {https://doi.org/10.1103/PhysRevD.66.066004} {\bibfield  {journal} {\bibinfo
  {journal} {Phys. Rev. D}\ }\textbf {\bibinfo {volume} {66}},\ \bibinfo
  {pages} {066004} (\bibinfo {year} {2002})},\ \Eprint
  {https://arxiv.org/abs/hep-ph/0205252} {arXiv:hep-ph/0205252} \BibitemShut
  {NoStop}%
\bibitem [{\citenamefont {Bachas}\ \emph {et~al.}(1996)\citenamefont {Bachas},
  \citenamefont {Fabre},\ and\ \citenamefont {Yanagida}}]{Bachas:1995yt}%
  \BibitemOpen
  \bibfield  {author} {\bibinfo {author} {\bibfnamefont {C.}~\bibnamefont
  {Bachas}}, \bibinfo {author} {\bibfnamefont {C.}~\bibnamefont {Fabre}},\ and\
  \bibinfo {author} {\bibfnamefont {T.}~\bibnamefont {Yanagida}},\ }\href
  {https://doi.org/10.1016/0370-2693(95)01561-2} {\bibfield  {journal}
  {\bibinfo  {journal} {Phys. Lett. B}\ }\textbf {\bibinfo {volume} {370}},\
  \bibinfo {pages} {49} (\bibinfo {year} {1996})},\ \Eprint
  {https://arxiv.org/abs/hep-th/9510094} {arXiv:hep-th/9510094} \BibitemShut
  {NoStop}%
\bibitem [{\citenamefont {Lopez}\ \emph {et~al.}(1996)\citenamefont {Lopez},
  \citenamefont {Nanopoulos},\ and\ \citenamefont {Zichichi}}]{Lopez:1996gd}%
  \BibitemOpen
  \bibfield  {author} {\bibinfo {author} {\bibfnamefont {J.~L.}\ \bibnamefont
  {Lopez}}, \bibinfo {author} {\bibfnamefont {D.~V.}\ \bibnamefont
  {Nanopoulos}},\ and\ \bibinfo {author} {\bibfnamefont {A.}~\bibnamefont
  {Zichichi}},\ }\href {https://doi.org/10.1103/PhysRevLett.77.5168} {\bibfield
   {journal} {\bibinfo  {journal} {Phys. Rev. Lett.}\ }\textbf {\bibinfo
  {volume} {77}},\ \bibinfo {pages} {5168} (\bibinfo {year} {1996})},\ \Eprint
  {https://arxiv.org/abs/hep-ph/9609524} {arXiv:hep-ph/9609524} \BibitemShut
  {NoStop}%
\bibitem [{\citenamefont {Barger}\ \emph {et~al.}(2005)\citenamefont {Barger},
  \citenamefont {Jiang}, \citenamefont {Langacker},\ and\ \citenamefont
  {Li}}]{Barger:2005qy}%
  \BibitemOpen
  \bibfield  {author} {\bibinfo {author} {\bibfnamefont {V.}~\bibnamefont
  {Barger}}, \bibinfo {author} {\bibfnamefont {J.}~\bibnamefont {Jiang}},
  \bibinfo {author} {\bibfnamefont {P.}~\bibnamefont {Langacker}},\ and\
  \bibinfo {author} {\bibfnamefont {T.}~\bibnamefont {Li}},\ }\href
  {https://doi.org/10.1016/j.nuclphysb.2005.08.007} {\bibfield  {journal}
  {\bibinfo  {journal} {Nucl. Phys. B}\ }\textbf {\bibinfo {volume} {726}},\
  \bibinfo {pages} {149} (\bibinfo {year} {2005})},\ \Eprint
  {https://arxiv.org/abs/hep-ph/0504093} {arXiv:hep-ph/0504093} \BibitemShut
  {NoStop}%
\bibitem [{\citenamefont {Jiang}\ \emph {et~al.}(2007)\citenamefont {Jiang},
  \citenamefont {Li},\ and\ \citenamefont {Nanopoulos}}]{Jiang:2006hf}%
  \BibitemOpen
  \bibfield  {author} {\bibinfo {author} {\bibfnamefont {J.}~\bibnamefont
  {Jiang}}, \bibinfo {author} {\bibfnamefont {T.}~\bibnamefont {Li}},\ and\
  \bibinfo {author} {\bibfnamefont {D.~V.}\ \bibnamefont {Nanopoulos}},\ }\href
  {https://doi.org/10.1016/j.nuclphysb.2007.02.025} {\bibfield  {journal}
  {\bibinfo  {journal} {Nucl. Phys. B}\ }\textbf {\bibinfo {volume} {772}},\
  \bibinfo {pages} {49} (\bibinfo {year} {2007})},\ \Eprint
  {https://arxiv.org/abs/hep-ph/0610054} {arXiv:hep-ph/0610054} \BibitemShut
  {NoStop}%
\bibitem [{\citenamefont {Barger}\ \emph {et~al.}(2008)\citenamefont {Barger},
  \citenamefont {Deshpande}, \citenamefont {Jiang}, \citenamefont {Langacker},\
  and\ \citenamefont {Li}}]{Barger:2007qb}%
  \BibitemOpen
  \bibfield  {author} {\bibinfo {author} {\bibfnamefont {V.}~\bibnamefont
  {Barger}}, \bibinfo {author} {\bibfnamefont {N.~G.}\ \bibnamefont
  {Deshpande}}, \bibinfo {author} {\bibfnamefont {J.}~\bibnamefont {Jiang}},
  \bibinfo {author} {\bibfnamefont {P.}~\bibnamefont {Langacker}},\ and\
  \bibinfo {author} {\bibfnamefont {T.}~\bibnamefont {Li}},\ }\href
  {https://doi.org/10.1016/j.nuclphysb.2007.10.013} {\bibfield  {journal}
  {\bibinfo  {journal} {Nucl. Phys. B}\ }\textbf {\bibinfo {volume} {793}},\
  \bibinfo {pages} {307} (\bibinfo {year} {2008})},\ \Eprint
  {https://arxiv.org/abs/hep-ph/0701136} {arXiv:hep-ph/0701136} \BibitemShut
  {NoStop}%
\bibitem [{\citenamefont {Jiang}\ \emph {et~al.}(2009)\citenamefont {Jiang},
  \citenamefont {Li}, \citenamefont {Nanopoulos},\ and\ \citenamefont
  {Xie}}]{Jiang:2008xrg}%
  \BibitemOpen
  \bibfield  {author} {\bibinfo {author} {\bibfnamefont {J.}~\bibnamefont
  {Jiang}}, \bibinfo {author} {\bibfnamefont {T.}~\bibnamefont {Li}}, \bibinfo
  {author} {\bibfnamefont {D.~V.}\ \bibnamefont {Nanopoulos}},\ and\ \bibinfo
  {author} {\bibfnamefont {D.}~\bibnamefont {Xie}},\ }\href
  {https://doi.org/10.1016/j.physletb.2009.05.012} {\bibfield  {journal}
  {\bibinfo  {journal} {Phys. Lett. B}\ }\textbf {\bibinfo {volume} {677}},\
  \bibinfo {pages} {322} (\bibinfo {year} {2009})},\ \Eprint
  {https://arxiv.org/abs/0811.2807} {arXiv:0811.2807 [hep-th]} \BibitemShut
  {NoStop}%
\bibitem [{\citenamefont {Jiang}\ \emph {et~al.}(2010)\citenamefont {Jiang},
  \citenamefont {Li}, \citenamefont {Nanopoulos},\ and\ \citenamefont
  {Xie}}]{Jiang:2009za}%
  \BibitemOpen
  \bibfield  {author} {\bibinfo {author} {\bibfnamefont {J.}~\bibnamefont
  {Jiang}}, \bibinfo {author} {\bibfnamefont {T.}~\bibnamefont {Li}}, \bibinfo
  {author} {\bibfnamefont {D.~V.}\ \bibnamefont {Nanopoulos}},\ and\ \bibinfo
  {author} {\bibfnamefont {D.}~\bibnamefont {Xie}},\ }\href
  {https://doi.org/10.1016/j.nuclphysb.2009.12.028} {\bibfield  {journal}
  {\bibinfo  {journal} {Nucl. Phys. B}\ }\textbf {\bibinfo {volume} {830}},\
  \bibinfo {pages} {195} (\bibinfo {year} {2010})},\ \Eprint
  {https://arxiv.org/abs/0905.3394} {arXiv:0905.3394 [hep-th]} \BibitemShut
  {NoStop}%
\bibitem [{\citenamefont {Kokorelis}(2016)}]{Kokorelis:2016ckp}%
  \BibitemOpen
  \bibfield  {author} {\bibinfo {author} {\bibfnamefont {C.}~\bibnamefont
  {Kokorelis}},\ }\href {https://doi.org/10.22323/1.263.0070} {\bibfield
  {journal} {\bibinfo  {journal} {PoS}\ }\textbf {\bibinfo {volume}
  {CORFU2015}},\ \bibinfo {pages} {070} (\bibinfo {year} {2016})},\ \Eprint
  {https://arxiv.org/abs/1610.01742} {arXiv:1610.01742 [hep-ph]} \BibitemShut
  {NoStop}%
\bibitem [{\citenamefont {Chen}\ \emph {et~al.}(2018)\citenamefont {Chen},
  \citenamefont {Gogoladze}, \citenamefont {Hu}, \citenamefont {Li},\ and\
  \citenamefont {Wu}}]{Chen:2017rpn}%
  \BibitemOpen
  \bibfield  {author} {\bibinfo {author} {\bibfnamefont {H.-Y.}\ \bibnamefont
  {Chen}}, \bibinfo {author} {\bibfnamefont {I.}~\bibnamefont {Gogoladze}},
  \bibinfo {author} {\bibfnamefont {S.}~\bibnamefont {Hu}}, \bibinfo {author}
  {\bibfnamefont {T.}~\bibnamefont {Li}},\ and\ \bibinfo {author}
  {\bibfnamefont {L.}~\bibnamefont {Wu}},\ }\href
  {https://doi.org/10.1140/epjc/s10052-017-5496-z} {\bibfield  {journal}
  {\bibinfo  {journal} {Eur. Phys. J. C}\ }\textbf {\bibinfo {volume} {78}},\
  \bibinfo {pages} {26} (\bibinfo {year} {2018})},\ \Eprint
  {https://arxiv.org/abs/1703.07542} {arXiv:1703.07542 [hep-ph]} \BibitemShut
  {NoStop}%
\bibitem [{\citenamefont {Chen}\ \emph {et~al.}(2020)\citenamefont {Chen},
  \citenamefont {Gogoladze}, \citenamefont {Hu}, \citenamefont {Li},\ and\
  \citenamefont {Wu}}]{Chen:2018ucf}%
  \BibitemOpen
  \bibfield  {author} {\bibinfo {author} {\bibfnamefont {H.-Y.}\ \bibnamefont
  {Chen}}, \bibinfo {author} {\bibfnamefont {I.}~\bibnamefont {Gogoladze}},
  \bibinfo {author} {\bibfnamefont {S.}~\bibnamefont {Hu}}, \bibinfo {author}
  {\bibfnamefont {T.}~\bibnamefont {Li}},\ and\ \bibinfo {author}
  {\bibfnamefont {L.}~\bibnamefont {Wu}},\ }\href
  {https://doi.org/10.1142/S0217751X20501171} {\bibfield  {journal} {\bibinfo
  {journal} {Int. J. Mod. Phys. A}\ }\textbf {\bibinfo {volume} {35}},\
  \bibinfo {pages} {2050117} (\bibinfo {year} {2020})},\ \Eprint
  {https://arxiv.org/abs/1805.00161} {arXiv:1805.00161 [hep-ph]} \BibitemShut
  {NoStop}%
\bibitem [{\citenamefont {Ellis}\ \emph {et~al.}(1991)\citenamefont {Ellis},
  \citenamefont {Kelley},\ and\ \citenamefont {Nanopoulos}}]{Ellis:1990wk}%
  \BibitemOpen
  \bibfield  {author} {\bibinfo {author} {\bibfnamefont {J.~R.}\ \bibnamefont
  {Ellis}}, \bibinfo {author} {\bibfnamefont {S.}~\bibnamefont {Kelley}},\ and\
  \bibinfo {author} {\bibfnamefont {D.~V.}\ \bibnamefont {Nanopoulos}},\ }\href
  {https://doi.org/10.1016/0370-2693(91)90980-5} {\bibfield  {journal}
  {\bibinfo  {journal} {Phys. Lett. B}\ }\textbf {\bibinfo {volume} {260}},\
  \bibinfo {pages} {131} (\bibinfo {year} {1991})}\BibitemShut {NoStop}%
\bibitem [{\citenamefont {Langacker}\ and\ \citenamefont
  {Luo}(1991)}]{Langacker:1991an}%
  \BibitemOpen
  \bibfield  {author} {\bibinfo {author} {\bibfnamefont {P.}~\bibnamefont
  {Langacker}}\ and\ \bibinfo {author} {\bibfnamefont {M.-x.}\ \bibnamefont
  {Luo}},\ }\href {https://doi.org/10.1103/PhysRevD.44.817} {\bibfield
  {journal} {\bibinfo  {journal} {Phys. Rev. D}\ }\textbf {\bibinfo {volume}
  {44}},\ \bibinfo {pages} {817} (\bibinfo {year} {1991})}\BibitemShut
  {NoStop}%
\bibitem [{\citenamefont {Amaldi}\ \emph {et~al.}(1991)\citenamefont {Amaldi},
  \citenamefont {de~Boer},\ and\ \citenamefont {Furstenau}}]{Amaldi:1991cn}%
  \BibitemOpen
  \bibfield  {author} {\bibinfo {author} {\bibfnamefont {U.}~\bibnamefont
  {Amaldi}}, \bibinfo {author} {\bibfnamefont {W.}~\bibnamefont {de~Boer}},\
  and\ \bibinfo {author} {\bibfnamefont {H.}~\bibnamefont {Furstenau}},\ }\href
  {https://doi.org/10.1016/0370-2693(91)91641-8} {\bibfield  {journal}
  {\bibinfo  {journal} {Phys. Lett. B}\ }\textbf {\bibinfo {volume} {260}},\
  \bibinfo {pages} {447} (\bibinfo {year} {1991})}\BibitemShut {NoStop}%
\bibitem [{\citenamefont {Blumenhagen}\ \emph
  {et~al.}(2007{\natexlab{b}})\citenamefont {Blumenhagen}, \citenamefont
  {Cvetic},\ and\ \citenamefont {Weigand}}]{Blumenhagen:2006xt}%
  \BibitemOpen
  \bibfield  {author} {\bibinfo {author} {\bibfnamefont {R.}~\bibnamefont
  {Blumenhagen}}, \bibinfo {author} {\bibfnamefont {M.}~\bibnamefont
  {Cvetic}},\ and\ \bibinfo {author} {\bibfnamefont {T.}~\bibnamefont
  {Weigand}},\ }\href {https://doi.org/10.1016/j.nuclphysb.2007.02.016}
  {\bibfield  {journal} {\bibinfo  {journal} {Nucl. Phys. B}\ }\textbf
  {\bibinfo {volume} {771}},\ \bibinfo {pages} {113} (\bibinfo {year}
  {2007}{\natexlab{b}})},\ \Eprint {https://arxiv.org/abs/hep-th/0609191}
  {arXiv:hep-th/0609191} \BibitemShut {NoStop}%
\bibitem [{\citenamefont {Haack}\ \emph {et~al.}(2007)\citenamefont {Haack},
  \citenamefont {Krefl}, \citenamefont {Lust}, \citenamefont {Van~Proeyen},\
  and\ \citenamefont {Zagermann}}]{Haack:2006cy}%
  \BibitemOpen
  \bibfield  {author} {\bibinfo {author} {\bibfnamefont {M.}~\bibnamefont
  {Haack}}, \bibinfo {author} {\bibfnamefont {D.}~\bibnamefont {Krefl}},
  \bibinfo {author} {\bibfnamefont {D.}~\bibnamefont {Lust}}, \bibinfo {author}
  {\bibfnamefont {A.}~\bibnamefont {Van~Proeyen}},\ and\ \bibinfo {author}
  {\bibfnamefont {M.}~\bibnamefont {Zagermann}},\ }\href
  {https://doi.org/10.1088/1126-6708/2007/01/078} {\bibfield  {journal}
  {\bibinfo  {journal} {JHEP}\ }\textbf {\bibinfo {volume} {01}},\ \bibinfo
  {pages} {078}},\ \Eprint {https://arxiv.org/abs/hep-th/0609211}
  {arXiv:hep-th/0609211} \BibitemShut {NoStop}%
\bibitem [{\citenamefont {Florea}\ \emph {et~al.}(2007)\citenamefont {Florea},
  \citenamefont {Kachru}, \citenamefont {McGreevy},\ and\ \citenamefont
  {Saulina}}]{Florea:2006si}%
  \BibitemOpen
  \bibfield  {author} {\bibinfo {author} {\bibfnamefont {B.}~\bibnamefont
  {Florea}}, \bibinfo {author} {\bibfnamefont {S.}~\bibnamefont {Kachru}},
  \bibinfo {author} {\bibfnamefont {J.}~\bibnamefont {McGreevy}},\ and\
  \bibinfo {author} {\bibfnamefont {N.}~\bibnamefont {Saulina}},\ }\href
  {https://doi.org/10.1088/1126-6708/2007/05/024} {\bibfield  {journal}
  {\bibinfo  {journal} {JHEP}\ }\textbf {\bibinfo {volume} {05}},\ \bibinfo
  {pages} {024}},\ \Eprint {https://arxiv.org/abs/hep-th/0610003}
  {arXiv:hep-th/0610003} \BibitemShut {NoStop}%
\bibitem [{\citenamefont {Li}\ \emph {et~al.}(2023)\citenamefont {Li},
  \citenamefont {Sun},\ and\ \citenamefont {Wu}}]{Li:2022cqk}%
  \BibitemOpen
  \bibfield  {author} {\bibinfo {author} {\bibfnamefont {T.}~\bibnamefont
  {Li}}, \bibinfo {author} {\bibfnamefont {R.}~\bibnamefont {Sun}},\ and\
  \bibinfo {author} {\bibfnamefont {L.}~\bibnamefont {Wu}},\ }\href
  {https://doi.org/10.1007/JHEP03(2023)210} {\bibfield  {journal} {\bibinfo
  {journal} {JHEP}\ }\textbf {\bibinfo {volume} {03}},\ \bibinfo {pages}
  {210}},\ \Eprint {https://arxiv.org/abs/2212.05875} {arXiv:2212.05875
  [hep-th]} \BibitemShut {NoStop}%
\bibitem [{\citenamefont {Mansha}\ \emph {et~al.}(2023)\citenamefont {Mansha},
  \citenamefont {Li},\ and\ \citenamefont {Wu}}]{Mansha:2023kwq}%
  \BibitemOpen
  \bibfield  {author} {\bibinfo {author} {\bibfnamefont {A.}~\bibnamefont
  {Mansha}}, \bibinfo {author} {\bibfnamefont {T.}~\bibnamefont {Li}},\ and\
  \bibinfo {author} {\bibfnamefont {L.}~\bibnamefont {Wu}},\ }\href
  {https://doi.org/10.1140/epjc/s10052-023-12167-6} {\bibfield  {journal}
  {\bibinfo  {journal} {Eur. Phys. J. C}\ }\textbf {\bibinfo {volume} {83}},\
  \bibinfo {pages} {1067} (\bibinfo {year} {2023})},\ \Eprint
  {https://arxiv.org/abs/2303.02864} {arXiv:2303.02864 [hep-th]} \BibitemShut
  {NoStop}%
\end{thebibliography}

\end{document}